\documentclass[preprint]{aastex}

\slugcomment{Accepted to the Astrophysical Journal, Jan. 23, 2017}

\shorttitle{Residual Gas \& Dust in 
Transition Objects and Weak T Tauri Stars}
\shortauthors{Doppmann et al.}

\newcommand{\Mdotstar}{\dot M_*}

\newcommand{\percc}{\rm \,cm^{-3}}

\newcommand{\persqcm}{\rm \,cm^{-2}}
\newcommand{\gpersqcm}{\rm \,g\,cm^{-2}}

\newcommand{\teff}{\,T_{\rm eff}}
\newcommand{\logg}{\,\log g}
\newcommand{\vsini}{v\sin i}
\newcommand{\mveil}{r_{\rm M}}

\newcommand{\kms}{{\rm km}\,{\rm s}^{-1}}

\newcommand{\Kband}{$K$-band}

\newcommand{\Mband}{$M$-band}

\begin{document}


\title{Residual Gas \& Dust Around 
Transition Objects and Weak T Tauri Stars
\footnote{Data presented herein were obtained at the W.M. 
Keck Observatory from telescope time allocated to the National 
Aeronautics and Space Administration through the agency's scientific 
partnership with the California Institute of Technology and the 
University of California.  The Observatory was made possible by the 
generous financial support of the W.M. Keck Foundation.}}


\author{Greg W. Doppmann\altaffilmark{2},
Joan R. Najita\altaffilmark{3}, and John S. Carr\altaffilmark{4}}

\email{gdoppmann@keck.hawaii.edu}
\email{najita@noao.edu}
\email{carr@nrl.navy.mil}


\altaffiltext{2}{W.M. Keck Observatory, 65-1120 Mamalahoa Hwy, Kamuela HI 96743}

\altaffiltext{3}{National Optical Astronomy Observatory, 950 N. Cherry Ave., Tucson, AZ 85719, USA}

\altaffiltext{4}{Naval Research Laboratory, Code 7213, Washington, DC 20375, USA}



\begin{abstract} 

Residual gas in disks around young stars can 
spin down stars, 
circularize the orbits of terrestrial planets, and 
whisk away the dusty debris that is expected to serve 
as a signpost of terrestrial planet formation.
We have carried out a sensitive search for residual gas and dust in the 
terrestrial planet region surrounding young stars  
ranging in age from a few Myr to $\sim 10$\,Myr in age. 
Using high resolution $4.7\micron$ spectra of transition objects and 
weak T Tauri stars,  
we searched for weak continuum excesses and CO fundamental emission, 
after making a careful correction for the stellar contribution to the 
observed spectrum. 
We find that the CO emission from transition objects is weaker 
and located further from the star than 
CO emission from non-transition T Tauri stars with similar stellar 
accretion rates.  
The difference is possibly the result of chemical and/or dynamical effects 
(i.e., a low CO abundance or close-in low-mass planets). 
The weak T Tauri stars show no CO fundamental emission down to low flux levels 
($5\times 10^{-20}-10^{-18}\, {\rm W\,m^{-2}}$).  
We illustrate how our results can be used to constrain the residual 
disk gas content in these systems and discuss their potential 
implications for star and planet formation.

\end{abstract}





\section{Introduction}

The circumstellar disks that surround stars at birth are believed
to dissipate through a combination of processes that include
accretion, planet formation, and photoevaporation \citep{alexander2014}.
In this way, stars evolve from accreting stars
surrounded by an optically thick disk of gas and dust 
that extends in to the stellar surface 
(a typical classical T Tauri star; CTTS)
to stars with little evidence for stellar accretion or a
surrounding dust disk (a weak T Tauri star; WTTS).
While most stars are CTTS at 1--2 Myr of age, 
almost all are WTTS beyond 5 Myr of age, 
based on the measured stellar accretion rates and 
dust continuum excesses associated with young stars 
\citep[e.g.,][]{mamajek2009,sicilia-aguilar2006,williams2011}.
The unusual spectral energy distributions (SEDs) of the population 
of young stars called
transition objects (TOs), which have a significant continuum excess at
long wavelengths ($> 10\micron$) and a weak excess at short wavelengths
($< 10\micron$), are thought to represent an intermediate,
transitional, phase of evolution between CTTS and WTTS.

In charting out disk evolution, the dust component is readily probed by 
continuum excesses, through the use of SEDs when the excess is strong
and through high resolution spectroscopy when the excess is weak
\citep[e.g.,][]{hartigan1989,hartigan1991,valenti1993}.
It is important to probe directly the lifetime of the gaseous
component, because the dust may not reliably trace the gas. 
A residual dust disk in the WTTS phase might be a signature of
accompanying residual gas.
Alternatively, it could be unrelated to the evolution of the gaseous
disk, e.g., if it is post-accretion debris produced by terrestrial 
planet formation or collisions of planetesimals leftover from 
the planet formation process 
\citep{kenyon2004,raymond2011,raymond2012},
i.e., similar to the situation for HD~21997 \citep{kospal2013} but at much smaller radii.
A residual gas disk might also persist in the absence of dust.

The extent and lifetime of residual gas disks is important for 
our understanding of planet formation and stellar rotation. 
Residual gas disks at AU distances may affect the outcome
of terrestrial planet formation by
damping the eccentricities of forming terrestrial planets or protoplanets
and/or inducing their radial migration \citep{kominami2002,agnor2002}.
If residual gas persists long enough to damp terrestrial planet 
eccentricities, it would help to explain the low eccentricities of
terrestrial planets in the solar system and surrounding other stars
\citep{hadden2014,shabram2016,vaneylen2015}.

Residual gas in the terrestrial planet region may help to 
reconcile the large discrepancy between the high occurrence rate of Earth-mass 
exoplanets ($\sim 20$\%) and the much lower incidence rate (few \%) 
of the warm debris that is the expected signpost of terrestrial planet 
formation at ages of $\sim 10$\,Myr. A residual gas disk with a 
column density 
$>10^{-5}$ of the minimum mass solar nebula (MMSN) that persists during 
the epoch of terrestrial planet assembly can quickly remove small 
grains from the terrestrial planet region or prevent their 
formation \citep{kenyon2016}. 
If either occurs, warm debris is not the reliable beacon of 
terrestrial planet formation that it is commonly regarded to be 
\citep{kenyon2004,raymond2011,raymond2012,leinhardt2015}. 

Residual gas within 0.5\,AU that persists into the WTTS phase could
also play a role in regulating the rotation rates
of young stars \citep[see][for a review]{bouvier2014} 
through the coupling of stars to their gaseous inner disks via  
stellar magnetic fields (``disk locking'').
\citet{gallet2013} have shown that the rotational periods
of stars as a function of age can be explained if solar mass stars
remain magnetically locked to their gaseous inner disks for
5\,Myr on average. 

Despite its potential significance for the outcome of star 
and planet formation, residual disk gas is more difficult 
to study than residual dust, 
especially with {\it in situ} gas diagnostics. As a result, 
stellar accretion rates are often used as a surrogate diagnostic.
The decline in stellar accretion with time 
\citep[e.g.,][]{sicilia-aguilar2006,sicilia-aguilar2010}
is consistent with the picture of a dissipating gaseous disk.
Surveys of both NIR excesses and stellar accretion find that 
fewer than 15\% of young stars show evidence of an inner disk at 5\,Myr
\citep[e.g.,][]{briceno2005,hernandez2005,fedele2010}. 
Stellar accretion rates would be a good probe of residual gas if we
could be certain that all of the gas in the disk is actively accreting. 
However, the reliability of this assumption is an open question.
It is therefore of interest to know whether 
gaseous inner disks survive for much longer than either 
the inner dust disk or gaseous accretion onto the star.

Here we explore this issue by carrying out a sensitive search for 
{\it in situ} emission from residual gas in disk systems with 
low stellar accretion rates.
We use high resolution $4.7\micron$ spectroscopy to search for
CO ro-vibrational emission which probes residual gas in the terrestrial
planet region of the disk.
Thermally excited CO requires high density ($>10^{10}\percc$) 
and temperature ($>500$\,K) to emit, conditions
that are characteristic of the inner disks of T~Tauri stars.
Detected from almost all CTTS, CO ro-vibrational emission 
probes disk radii from the inner disk edge to $\sim 1$\,AU for typical CTTS, 
based on spectroscopically resolved CO emission.
These properties make it an attractive probe of gaseous disks
in the terrestrial planet region \citep{najita2003,najita2007,brittain2003,
brittain2007,blake2004,rettig2004,salyk2008,salyk2009,salyk2011,brown2013,banzatti2015}.

CO emission from transition objects has been studied 
previously \citep{najita2003,najita2008,rettig2004,salyk2007,salyk2009,pontoppidan2008}, 
although similar studies have not been carried out for WTTS. 
Because the strength of the emission tends to decline with
decreasing accretion rate \citep{najita2003,brittain2007},
it is important in studying WTTS to optimize for the detection of 
potentially weak CO emission.
To accomplish this, in this study we correct for stellar absorption features in
the composite (star + disk) spectra of our sample.

With this method, we are also sensitive to 
weak continuum excesses (continuum veiling) which are a measure of
warm dust close to the star.
Although veiling is well studied at shorter wavelengths,
the $5\micron$ region has received less attention
\citep{najita2008,salyk2009}. 
The use of CO as a probe of residual gas in TOs and WTTS complements
other related studies that use mid-infrared H$_2$ lines \citep[e.g.,][]{pascucci2006}, 
UV H$_2$ features \citep{ingleby2009,ingleby2012,france2012},
and other mid-infrared diagnostics \citep{pascucci2007,najita2010}
as {\it in situ} tracers of residual gas.

Our \Mband\ spectroscopic observations and data reduction are described in $\S$2.  The modeling technique and analysis is described in $\S$3, and the results reported in $\S$4.  We discuss our findings in $\S$5 and conclusions in $\S$6.


\section{Observations} 

For our study of the dissipation of gas and dust in the 
inner disks of TOs and WTTS, we selected a small 
sample of well characterized TOs and WTTS in Taurus
and the TW Hydra Association (TWA). The average age of the sources in 
these regions (few Myr for Taurus and $\sim 10$\,Myr for TWA) 
spans the expected gas dissipation timescale.  
Because our gas diagnostic, CO ro-vibrational emission, is known to 
decline in strength with stellar accretion rate 
at high $\Mdotstar$ \citep{najita2003},
we also studied a small sample of CTTS with low $\Mdotstar$ 
in order to determine if CO remains a reliable gas diagnostic 
at low $\Mdotstar$.
Our sample (Table 1) includes 4 CTTS with 
relatively low accretion rates 
\citep[$\sim 10^{-9}$M$_{\odot}$yr$^{-1}$; 
IP~Tau, DN~Tau, V836~Tau, ZZ~Tau,][]
{najita2015,valenti1993}, 5 well-studied 
transition objects \citep[TW~Hya, UX~Tau~A, LkCa~15, GM~Aur, CoKu~Tau/4,][]{furlan2006}  
and 15 WTTS \citep{furlan2006,furlan2011}.
The targets span a range in spectral type (K0-M3; Table 1). 
The majority of the sample are apparently single stars. 
ZZ~Tau and Coku~Tau/4 are 0.06 arcsec binaries 
\citep{schaefer2006,ireland2008}.
Hubble~4, HBC~427, V827~Tau, and TWA~5 are spectroscopic binaries
\citep{walter1988,neuhauser2001,kraus2011,mahmud2011}.

The data were
obtained over multiple observing runs from 2001 January, 2003
January, 2005 February, and 2007 January (Table 1).
We took \Mband\ spectra of the Taurus and TWA objects in our sample 
using the Keck II facility
spectrometer, NIRSPEC \citep{mclean1998} at the summit of Mauna Kea.  
We obtained multi-order spectra at high resolution (R=25,000 and
R=18,500), with echelle and cross-disperser gratings oriented to
give spectral coverage spanning $4.586-4.662\micron$ (order 16)
and $4.890-4.972\micron$ (order 15).  Depending on the seeing
conditions, the 0.432$\arcsec$ $\times$ 24$\arcsec$ slit or
0.576$\arcsec$ $\times$ 24$\arcsec$ slit was used to acquire spectra
through the M-wide broadband filter, imaging echelle orders 15, 16
and 17 onto the 1024 $\times$ 1024 Aladdin-3 detector.

The large thermal sky background required short exposure times ($1-2$ sec) for all
our targets, and were co-added
internally ($20-60$ times).  The telescope was nodded $\pm$6
$\arcsec$ along the slit in an ABBA pattern to obtain
simultaneous source and sky spectra at each nod position.   Spectra
of early type standards were taken at similar airmass for removal
of telluric absorption lines in the targets.  Spectra of flux
standards were taken through the widest available slit (5 pixels,
0.720$\arcsec$) for spectroscopic flux calibration.
High signal-to-noise flat fields for the \Mband\ were obtained by
combining multiple (i.e., 20) continuum lamp images of short exposure
time (1 sec), taken with the NIRSPEC internal calibration unit.

The data were reduced by Nathan Crockett following the procedure
described in $\S 2$ of \citet{najita2008} using IRAF packages
\citep{massey1992,massey1997}.  Exposure frames from nod pairs were
differenced then divided by a normalized flat field image.  Using
spatially resolved telluric emission sky lines present in each exposure, the images
were rectified  to create mono-spectral columns and mono-spatial
rows, which allowed a large background aperture to be used when
extracting the spectral trace of the source, producing a higher
signal-to-noise spectrum.  Rectified images from each beam position
were shifted, as needed, to place the peak of each source aperture
on the same pixel before co-adding them.

Spectra of target and telluric calibration stars were extracted
from an aperture profile having a width of $\pm$ 2 pixels ($\pm~
0.39\arcsec$) about the spatial peak and subtracted by a background
aperture that was more than 7 pixels ($1.35\arcsec$) from the source
peak on both sides, spanning a total width of $\sim$40 pixels
($\sim$7.8$\arcsec$).  Telluric absorption features were removed
by dividing extracted target spectra by the normalized telluric spectra taken
at similar airmass.

Our CTTS and TO target spectra
were flux calibrated from corresponding telluric
star observations that were taken close in time.  Except in three
cases (i.e., UX~Tau~A in 2005, V836~Tau in 2003, and IP~Tau), wide slit (i.e.,
$0.72\arcsec$) observations were unavailable or were of insufficient
signal-to-noise.  Instead, narrower slit observations (i.e.,
$0.43\arcsec$) for both target and telluric standard were used for
flux calibration under the assumption of equal slit losses.  In
some cases, a comparison of summed counts between different nod
positions within a given target or telluric observation sequence
differed by more than $15\%$. 
Consequently, the errors in the flux
calibration of GM~Aur, DN~Tau, TW~Hya (in 2001), IP~Tau, V836~Tau
(in 2003), and LkCa~15 (in 2001) are dominated by uncertainty in
the flux throughput caused by variable slit losses in the target
and telluric observations.

To correct for the broad Pf$\beta$ absorption at $4.650\micron$ in
our early-type telluric standards, we created a spline fit to the
Pf$\beta$ feature in the spectrum of the telluric standard and 
then multiplied the fit into the (telluric-corrected)
target spectrum to restore its true continuum shape (Fig.\ 1).


\section{Analysis}

Weak fundamental CO line emission from the warm inner disk 
can be obscured in the composite (star + disk) 
spectrum by CO absorption lines from the stellar photosphere.  
Thus we need a template of the pre-main-sequence stellar photosphere 
in order to accomplish our goals of 
(i) detecting and measuring potentially weak CO emission from the disk and 
(ii) measuring the excess emission over the stellar photosphere, 
i.e., the veiling at \Mband\ ($\mveil$). 
Because it is difficult to obtain a complete set of observational templates 
(i.e., high signal-to-noise spectra of disk-less PMS stars that span 
the range in $\teff$, $\logg$, and $v\sin i$) in our sample, 
we created synthetic spectral 
templates with the appropriate properties for each source. 

\subsection{Spectral Fitting Method}

We used MOOG \citep{sneden1973} to generate synthetic \Mband\ stellar
spectra for these sources following the procedure detailed in \citet{najita2008}.
The inputs to MOOG are a model atmosphere for a given effective temperature
and gravity,
rotational and instrumental broadening, microturbulance, and a limb
darkening coefficient.
We used the Allard NG 5.0 model atmospheres \citep{hauschildt1999}, 
formatted to run as input to MOOG.
We also assumed solar abundances, consistent with stellar abundance measurements of
nearby star forming regions \citep{padgett1996,santos2008},
and stellar rotational velocity broadening values from the literature
(Table 1), convolved with the instrumental broadening of our
slit resolution. 
We adopted 
a limb darkening coefficient of 0.6, a microturbulence of 0.5 $\kms,$ 
and theoretically calculated oscillator strengths 
for $^{12}$CO and $^{13}$CO transitions \citep{goorvitch1994}
to generate model CO stellar absorption spectra.

Effective temperature ($\teff$) and surface gravity ($\logg$) are
the primary input parameters for the spectral synthesis.  
Because veiling and $\logg$ have a similar effect on our $4.7\micron$ 
spectra,  we were unable to measure $\logg$ from our spectra themselves. 
We therefore used literature values to determine the input $\logg$ 
and $\teff$ (Table 1) and then fit for veiling.
The uncertainties in the resulting effective temperatures
are typically $\Delta \teff \sim 150K$ ($\sim$ 1~spectral subclass).  
At a fixed gravity of $\logg = 4.0$,
which is typical for our sample, 
the uncertainty in $\teff$ 
corresponds to a typical uncertainty 
in the veiling of $\Delta \mveil  \sim 0.05$.
For most of our sources, we estimated $\logg$ using stellar luminosities and temperatures or spectral types from the literature,
and theoretical evolutionary model tracks (Table 1).  

We adopted the spectral types, luminosities, and masses of \citet{herczeg2014} with a few exceptions. 
For V410~Tau, which was not included in that study, we used the effective temperature and luminosity from \citet{kenyon1995} and \citet{bertout2007}. The luminosity was converted to gravity using the stellar mass derived from the model tracks of \citet{baraffe2015}. We also adopted 
$\teff$ and $\logg$ values from spectroscopic fits when available
\citep[][UX Tau A, LkCa 19, and Hubble 4]{balachandran1994, johns-krull2004}. 
Errors on the luminosity and derived temperature values corresponded to a typical range of $\pm 0.15$ in $\logg$.
Using the Siess et al. (2000) tracks instead would result in lower gravities by $\sim 0.05$ in the log.

For TW Hya, we adopted the temperature and gravity reported by \citet{yang2005} from detailed 
modeling of high resolution, high signal-to-noise spectra. For TWA 10, we adopted the 
spectral type from \citet{webb1999} and estimated its gravity from the bolometric luminosity of 
\citet{delareza2004} and the mass predicted by  \citet[][Table 1]{baraffe2015}.

Uncertainties in the photometry, distance, and position of theoretical
tracks yielded an uncertainty in the inferred gravity of $\Delta
\logg \sim \pm 0.35$.  
Across the temperature range of ($\teff$ =3400--4400K), 
characteristic of many of our sources, the uncertainty
in the surface gravity gives an uncertainty of order ($\Delta \mveil
\sim \pm 0.15$) in the veiling.   Generally, the increasing
strength of the CO absorption lines in synthesis models with
decreasing temperature and/or decreasing gravity is offset with
increased veiling.

Since the input model atmosphere structures for our synthetic
spectral modeling were more coarsely gridded in $\teff$ ($\Delta
\teff = 100\rm K$ for $\teff < 4000\rm K$, and 
$\Delta \teff = 200\rm K$ for $\teff \ge 4000\rm K$) 
and $\logg$ ($\Delta \logg = 0.5$) than the values
we used from the literature, we chose the nearest model value in
each case (see Table 2).  
Rotational broadening and radial velocity could be determined 
directly from our model fits, and our observations confirmed values 
reported in the literature, except in a few cases (Table 2).
For each source, we used the $\vsini$
velocity broadening value from the literature 
\citep{nguyen2012,yang2005,yang2008,johns-krull2004,white2004,
delareza2004,reid2003,torres2003}
in the model fits 
unless a better fit value could be obtained 
(Table 1). 

To validate the models, we first fit the spectra of two main sequence 
dwarf standards with well-known stellar parameters (i.e., effective 
temperature, surface gravity, abundances) that were observed
with the same instrumental setup as our sources.  We found a
good fit to the observed spectra (Fig.~2) using a model with the known 
stellar parameters of 61~Cyg A and B: 
K5V and K7V spectral types respectively, 
a gravity of $\logg = 4.5$, which is appropriate for a dwarf luminosity
class of a late-type main sequence star and is consistent with interferometric
radius measurements \citep{kervella2008}, 
and a metallicity of [Fe/H]=-0.27 \citep{luck2005}.
As we expected for these normal main sequence stars, we found the
best fit veiling value to be consistent with zero, after testing for negative and
positive values of veiling.

We also fit the spectrum of a late-type
dwarf standard (M0V) observed in a nearby spectral region \citep[HD~79211,
Salyk private communication; see also][]{salyk2009}, using a
model with a temperature ($\teff=3800K$) and gravity ($\logg=5.0$)
expected for this main sequence star.  We find that the model fits
the observed standard very well with zero veiling, as expected.

\subsection{Measured Veiling and CO Emission}

We determined the \Mband\ continuum veiling of our sources by 
fitting their observed spectra with a sum of an appropriate 
synthetic stellar template ($\S 3.1$) and a featureless veiling continuum 
with a strength $\mveil$ times the stellar continuum.  
The best-fitting value of $\mveil$ was determined using the RMS pixel-to-pixel fit
in selected sub-regions in our spectra that had telluric transmission 
$\ge 80\%$ 
and were free of both artifacts and circumstellar CO and Pf$\beta$ emission. 
Veiling values $\mveil \simeq$ 0--5 were obtained (Table 2).  
To estimate the uncertainty in $\mveil$ due to the 
uncertainty in $\teff$ ($\pm 150K$) and $\logg$ ($\pm 0.25$), 
we considered pairs of $\teff$ and $\logg$ values within the 
allowed range that maximized or minimized the strength of 
the stellar absorption lines in our templates and 
fit for the corresponding values of $\mveil$. 
We adopted the range in the fit values as our uncertainty in $\mveil$ 
(Table 2). 
Generally, 
the WTTS in our sample show little or no veiling (e.g., Fig. 3a), 
while the CTTS have significant \Mband\ veiling (e.g., Fig 3c).

We expect the veiling to be underestimated if one uses a
main sequence stellar template to fit pre-main sequence sources.
To illustrate this point, we can look at possible model fits to 
CoKu~Tau/4, a source with a gravity that is 
typical of the pre-main-sequence sources in our sample, $\logg \sim 4.2$.
Because this value falls between the $\logg$ values sampled 
by our atmosphere models ($\logg=$4.0, 4.5, and 5.0), 
we adopted a model with the closest gravity $\logg=4.0$, for which 
the best fit veiling is $\rm=0.3$ (Fig.~3b).\footnote{While we could 
interpolate between model fits to obtain a more precise value 
(i.e., since the $\logg=4.5$ model requires a veiling of $\rm=0.1$, 
the veiling for $\logg=4.2$ is close to $r_M=0.2$), it 
would not alter the results reported here. As a result, we report 
instead the properties inferred using the model with the closest 
$\logg$ and $\teff$.} 
If we had used a model with main sequence gravity 
(i.e, $\logg=5.0$ and $\teff=3800K$)
or an equivalent main sequence star as a template 
\citep[e.g., HD~79211][]{salyk2009}, 
we would have inferred a lower veiling value of $r_M=-0.1$.  
Thus, using
spectral synthesis templates enables a closer match in $\logg$ and
$\teff$, enabling 
the identification of weak continuum veiling.

Although weak CO emission can be obscured by stellar CO absorption
features, it can be unveiled by correcting for the veiled stellar continuum. 
To search for weak CO emission, 
we subtracted the synthetic
veiled stellar spectrum and replaced it with a featureless continuum
of the same strength.  We then measured any detected CO and Pf$\beta$
emission. 

Table 3 reports measured CO emission equivalent widths and 
1-$\sigma$ upper limits 
relative to the total (stellar + veiling) continuum. 
The upper limits include contributions from pixel-to-pixel noise and the 
uncertainty in the determination of the continuum level.  
The CO equivalent widths reported in Table 3 were multiplied by the 
continuum flux at the middle of the corresponding order 
($4.63\micron$ or $4.93\micron$) to obtain CO emission fluxes. 
For the WTTS, whose spectra were not flux calibrated ($\S$2), 
we first estimated their flux at the center of the 
\Mband\ from their \Kband\ magnitudes in the literature and 
assuming main sequence $K$--$M$ colors \citep{kenyon1995} 
appropriate for their spectral type. 
We then used the spectral shape of IP~Tau through the \Mband\ as 
a template to estimate the WTTS fluxes at $4.6\micron$ and $4.9\micron.$ 
The spectral shape for IP~Tau was obtained from 
the flux calibrated measurements reported here, which had appropriate
wide slit observations at high signal-to-noise.   


\section{Results}

We detect 1-0 $^{12}$CO emission from 7 of the 24 sources in our sample (Fig. 4, Table 3):
all of the non-binary, low accretion rate CTTS (DN~Tau, IP~Tau, V836~Tau) and  
all of the non-binary TOs (GM~Aur, LkCa~15, TW~Hya, UX~Tau~A). 
For the detected CO emission sources, 
we constructed average CO line profiles (Fig.~5) for both R-branch
(red) and P-branch (blue) lines by interpolating 
individual lines onto a common velocity grid and averaging the 
emission, excluding regions of poor telluric correction 
or other artifacts. 
Table 3 lists the lines that
were included in the average profiles along with their resulting
FWHM and equivalent widths.  
When CO emission is detected, the emission profiles
are symmetric and centered within 4 $\kms$ of the stellar velocity, as reported in the literature. (Fig.~5).

We also detect Pf$\beta$ emission in 2 sources (TW~Hya, GM~Aur), 
with 4 others (LkCa~15, UX~Tau~A, IP~Tau, DN~Tau)
showing marginal detections (Fig.\ 6).
The Pf$\beta$ emission detected from GM~Aur and TW~Hya in 2001 and 
2003 had equivalent widths of 24\AA, 34\AA, and 63\AA, 
and FWHM of $206\,\kms,$ $99\,\kms,$ and $95\,\kms,$ respectively. 

Assuming that the CO emission arises in a Keplerian disk 
\citep{salyk2011,brittain2007,najita2003,banzatti2015},
the broad line profiles of the low accretion rate CTTS (K7/M0; 120--150 $\kms$ FWHM)
indicate that most of the CO emission from these sources 
arises close to the star, within 0.5\,AU, given their system 
inclinations \citep{andrews2007,ardila2013,najita2008}
and stellar masses \citep{andrews2013} from the literature. 
The narrower line profiles of the CO emission 
from the TOs (10--80 $\kms$ FWHM, deconvolved) 
imply that their CO emission arises from larger radii (0.3--2\,AU) 
given their higher average masses 
\citep[corresponding to their K2-K7 spectral types;][]{andrews2013}
and system inclinations 
\citep{andrews2011a,pontoppidan2008}.

\citet{salyk2009} reached the same conclusion in their study, 
which included the same TOs studied here \citep[see also][]{banzatti2015}. 
Similar in spirit to these results, \citet{hoadley2015} found that the line profiles of 
UV fluorescent H$_2$ emission from TOs is narrower than those of classical T Tauri stars, 
with the narrower profiles implying inner emission radii $\sim$4 times larger on average than for classical T Tauri stars.
We confirm that this conclusion also holds when the CO line widths of TOs are compared 
to the CO line widths of CTTS with similar accretion rates.  A similar trend has also been found for the [OI] emission width from TOs. In the study of \citet{simon2016}, three of the four TOs studied lack a broad [OI] low velocity component, suggesting that the emission arises from gas further away from the star than the [OI] emission from CTTS.

The inferred CO emission radii for the TOs are well within the sub-millimeter 
cavity for these sources
\citep[radii of 2\,AU, 25\,AU, 28\,AU, 50\,AU
for TW~Hya, UX~Tau~A, GM~Aur, and LkCa~15, respectively; ][]
{andrews2016,andrews2011b}.
The presence of gas within the sub-millimeter cavity
is consistent with the results of previous studies of CO emission 
from transition objects \citep{rettig2004,najita2008,salyk2007,salyk2009,
pontoppidan2008}
and the ongoing stellar accretion in these systems, which also 
indicates gas close to the star.

As shown in Tables 2 and 3, all of the sources from which CO emission 
is detected have 
significant veiling ($r_M > 0.5$) and measurable accretion.
CO emission is not
detected from WTTS and non-accreting transition objects (CoKu~Tau/4).
Our results are similar to those of \cite{salyk2009, salyk2011} who found
that CO emission is detected from some but not all transition objects.
In several sources we detect apparent non-zero \Mband\ veiling 
($\mveil \ge 0.3$) but no CO emission.
One of these sources, 
TWA~7, also has a detected continuum excess at longer wavelengths with 
{\it Spitzer} \citep{low2005}. The lack of accompanying tracers of a 
gas disk is consistent with the suggestion that the continuum excess 
arises from collisional debris produced through planet building. 
Although the WTTS V819~Tau also has a reported excess at 
{\it Spitzer} wavelengths 
\citep{furlan2009}, we did not measure an \Mband\ excess for 
this source (Table 2). 
The TWA 2AB binary, which is spatially unresolved in our spectrum,
has components with similar spectral types (M1.7 and M3.5 for A and
B, respectively) and a luminosity ratio of 3.4 
\citep{herczeg2014}. If all of the veiling
is associated with the A component, the veiling we measure ($r_M=0.4$)
only slightly underestimates the veiling of the A component ($r_M=0.5$).

We measure \Mband\ veilings of $r_M = -0.2$ to 0.4 for our WTTS sources and 
$r_M=0.3$ to 5.1 for transition objects (Table 2). 
Negative veiling values may result if the literature values of 
$\teff$ and $\logg$ are not quite right and/or from the finite 
gridding in $\teff$ and $\logg$ of the model atmospheres.
The difference between our inferred veiling for CoKu~Tau/4 of $r_M=0.3$ 
and the smaller value from \cite{salyk2009} 
results from the lower gravity stellar template used in our study. 
The lower gravity of CoKu~Tau/4 increases 
the strength of its stellar CO absorption lines 
compared to a main sequence star of the same spectral type, which is 
compensated for by modest ($r_M=0.3$) continuum veiling. 
As a result, the strength of the CO absorption features are similar for 
the main sequence standard and 
the (veiled) spectrum of CoKu~Tau/4. 
Future observations that enable more precise values of $\logg$ 
(e.g., more precise distance estimates) will allow us to 
determine $r_M$ more precisely.

Figure 7 plots $r_M$
against the $24\micron$/$K_s$ flux ratio for our sample \citep{low2005,
rebull2010,furlan2006,furlan2011}. 
Transition objects with large $24\micron$/$K_s$ flux ratios (1--6) 
have a range of $r_M$ (0.3--5), whereas WTTS are clustered toward
small excesses on both axes. The low accretion rate T Tauri stars
in our sample have intermediate $24\micron$/$K_s$ colors (0.4--1) 
and span the same range of $r_M$ as the TOs. 
The high $r_M$ for the transition objects LkCa~15 and UX~Tau~A is
consistent with their classification as ``pre-transition disks'', 
systems with transition disk SEDs that also have a significant 
near-infrared excess indicative of an optically thick inner disk 
\citep{espaillat2010}. 

For objects studied at 2 epochs, we can compare any variation in 
the CO flux and \Mband\ veiling flux, which were measured simultaneously. 
Figure 8 plots the measured CO equivalent widths against $r_M$ 
(Tables 2 and 3) for all sources. Because the CO equivalent width 
is multiplied by $(1+r_M)$, this product represents the CO emission flux
ratioed to the \Mband\ stellar continuum flux.
Thus the two axes show the continuum excess and the CO emission flux
relative to the stellar continuum.
Dashed lines connect sources observed at more than one epoch.
For LkCa~15 and TW~Hya, which have multiple measurements, we find 
an increase (decrease) in the continuum excess is accompanied by 
an increase (decrease) in the CO emission flux. 
These variations could result from increased accretion, which can increase 
the column density of the gas and dust in the inner disk and/or raise the 
temperature of the emitting gas. 
However, for two other sources, a change in $r_M$ (UX~Tau~A) or 
CO emission strength (V836~Tau) was not accompanied by a significant
change in the other quantity. 
Figure 8 also shows that some transition objects can have higher \Mband\ veiling 
than the low accretion rate T Tauri stars (e.g., LkCa~15), 
and that the transition objects have systematically lower CO emission flux 
than low accretion rate T Tauri stars with the same \Mband\ veiling.

Figure 9 compares measured CO emission fluxes with stellar
accretion rates from the literature, for both the sample studied here
(large symbols) 
and sources from the literature (small symbols) 
\citep{brown2013,najita2003,najita2007,
natta2006,valenti1993,espaillat2010,eisner2005}. 
The CO fluxes are normalized to a distance
of 140\,pc, assuming distances of 140\,pc for Taurus sources,
120\,pc for Oph sources, and 56\,pc for TWA sources.
CO emission strength generally increases with stellar accretion
\citep[see also][]{najita2003,brittain2003},
although there is significant dispersion.
The transition objects studied here cluster in the lower left
region of the plot, i.e., they have low CO emission fluxes for their
stellar accretion rates.

WTTS show no CO emission down to equivalent widths of 
$\sim$0.1\AA~and fluxes of 10$^{-19}$ W~m$^{-2}$.
The upper limit on the CO emission flux for Hubble~4 
is particularly large because it is very bright. 
The WTTS accretion rate upper limits shown in Figure 9 
are fiducial limits based on the work of \citet{manara2013}
who estimated the limit on measurable stellar accretion that is 
imposed by typical levels of stellar chromospheric emission. 
The limiting values depend on stellar mass and age.  Thus 
we use individual masses, as derived from the gravities (Table 1), and representative
ages for TWA (i.e., 10 Myr)  and Taurus (i.e., 2 Myr) to estimate the accretion rate upper limits for the
WTTS in our sample.


\section{Discussion}

\subsection{CO Emission and \Mband\ Excess from Transition Objects}

We find that CO ro-vibrational emission generally declines with decreasing 
stellar accretion rate (Fig.\ 9). CO emission 
also declines from CTTS to TO to WTTS classes, with an overlapping
spread in emission strength in the CTTS and TO classes. 
Transition objects have lower CO emission flux than 
TTS with 
similar stellar accretion rate (Fig.\ 9) or 
similar \Mband\ veiling (Fig.\ 8).  
They also have narrower CO emission line profiles than 
CTTS with similar accretion rates and inclinations 
\citep[see also][]{salyk2009}.

Although we generally expect gas and dust diagnostics to be related, 
the lack of a strong correlation between $r_M$ and CO emission
flux in our sample (Fig.~8) is not all that surprising, 
because $r_M$ and CO emission are affected by distinct processes. 
While \Mband\ veiling should increase linearly with dust
mass when the dust disk is optically thin, other effects 
enter when the dust is optically thick. Dust scale heights 
can vary from source to source depending on the gas temperature 
and the extent of grain settling out of the atmosphere, and 
shadowing effects can come into play \citep[e.g.,][]{espaillat2011}.  
In parallel, CO emission strength can depend on factors other than 
total gas mass, e.g., gas temperature, CO abundance, and optical depth.
The scatter seen in Figure 8 could be due to differences in 
some of these same factors
among sources with similar accretion rates. 

Interestingly, some of the transition objects are mixed in with the 
CTTS in Figure 9. SR24~S and DoAr~44 (open gray circles), 
both located in Ophiuchus, 
have some unusual properties as transition objects. 
SR24~S is identified as transition object based on the
inner cavity in its sub-millimeter continuum emission 
\citep{andrews2011b} rather than by its SED. 
Its SED is unremarkable and is consistent with
an inner disk that is optically thick in the near- and mid-infrared. 
DoAr~44 has a more typical transition object SED, with a 
mid-infrared ``dip'' \citep{espaillat2010}, 
although it is 
unusual in that it has a rich molecular emission spectrum at 
{\it Spitzer} wavelengths \citep{salyk2015}. 
The strong CO emission from these two sources is consistent with the 
evidence for substantial small dust grains (SR24~S) and
substantial gas (DoAr~44) close to the star.

In contrast to these two sources,
the CO emission from the rest of our TO sample is weaker than that from
CTTS of the same $r_M$ or stellar accretion rate 
(Fig.~8, open triangles in Fig.~9).
The reduced CO emission strength of TOs compared to CTTS could 
result from a reduction in the 
CO abundance, 
emitting area, or   
temperature of the gaseous disk,  
or some combination of these. 

\noindent{\bf Reduced CO abundance.}\quad
The weaker mid-infrared excess of TOs compared to CTTS 
has been interpreted as evidence for a reduced grain abundance 
in their inner disks, possibly as a consequence of planet formation. 
The accumulation of grains into larger objects (such as 
rocks, planetesimals, and protoplanets) 
will reduce the grain abundance in the inner disk. 
A similar result could occur when 
a giant planet forms and opens a gap at large disk radii
(beyond the CO emission region), and the resulting pressure bump at the 
outer edge of the gap filters out grains from the material 
reaching the inner disk \citep{rice2006,zhu2012}.
In disks with a highly reduced grain abundance, CO may form less 
efficiently because the formation of the precursor molecule H$_2$ 
on grain surfaces is less efficient (see section 5.2). 
The resulting CO abundance may be suppressed more strongly close to the star,  
for a given grain reduction factor, because of the higher rate of 
photodestruction at smaller disk radii. This effect would favor 
CO emission from larger disk radii, resulting in smaller CO emitting 
areas and narrower CO line profiles \citep[as in Fig.~5 and][]{salyk2009}.

\noindent{\bf Reduced gas emitting area.}\quad
Another possibility is that the weaker CO emission from TOs 
reflects a smaller emitting area for the gaseous inner disk as whole, 
not just the radial extent of the region where CO is abundant.  
In the \citet{simon2016} study of [OI] emission from TTS, 
the FWHMs of the [OI] and CO lines are very similar for the 3 TOs 
where both lines are detected (GM~Aur,UX~Tau4, TW~Hya). 
The similar [OI] and CO line profiles may suggest that both 
the atomic and molecular gas are similarly restricted to larger 
radii than in CTTS disks, as in a truncated gaseous disk.
Because TOs have significant stellar accretion rates, gas from the 
disk clearly reaches the star. 
A giant planet located close to the star can truncate the inner disk 
and concentrate the gas in accretion streams, reducing the gas emitting area. 
Because the CO emission from CTTS arises close to the star 
\citep[$\lesssim$ 0.3 \,AU;][]{salyk2011,najita2003}, 
a planet would likely have to be located 
within an AU to impact the CO-emitting region of the disk.
The narrower CO emission profiles of TOs, 
compared to those of CTTS with similar accretion rates and inclinations,
suggests that any truncation of the CO emission occurs from within 
(rather than beyond) the emission region, i.e., it requires a 
planet at $\sim 0.1$\,AU, similar to the close-in giant planet 
that is reported to orbit V830~Tau \citep{donati2016}. 

Any planet capable of affecting the structure of the gas 
in the CO-emitting region would be distinct from the giant 
planets that could create the much larger inner holes inferred 
for the TOs in our sample, 
i.e., TOs would be multiple planet systems \citep{zhu2011,dodson-robinson2011}. 
The planet that creates the inner hole would be located 
far from the CO emission region, near the  
inner hole radius 
\citep[2\,AU, 28\,AU, 25\,AU, 50\,AU for 
TW~Hya, GM Aur, UX Tau A, and LkCa 15;][] 
{andrews2011b,andrews2016}, 
as imaged directly in one case 
\citep[LkCa 15b at 14.7\,AU;][]{kraus2012,sallum2015}.

If a high mass (giant) planet is needed to truncate the disk 
close to the star, 
such planets are rarer among mature sun-like stars 
than TOs are among young stars. 
Hot Jupiters ($P<10$\,d; $M\sin i > 0.1 M_J$)  are present in 
$\sim 1$\% of sun-like stars \citep{wright2012}, 
whereas TOs represent 10--15\% of the disk-bearing 
population of young stars 
\citep{muzerolle2010,espaillat2014}. 
The low incidence rate of hot Jupiters around mature 
sun-like stars seems to argue against this interpretation. 

This scenario is more plausible if the same dynamical effect can be 
achieved with lower mass planets, which are more common as 
close companions to sun-like stars. 
Along these lines, \citep{andrews2016} have suggested that 
a young super-Earth (several $M_E$) may be responsible for the 
narrow gap observed in the inner 1~AU of the TW~Hya disk.

\noindent{\bf Reduced gas temperature.}\quad
Alternatively, TOs may show weak CO emission as a result of reduced 
gas temperature in the inner disk \citep[see also][]{banzatti2015}. 
The lower accretion rates of TOs lead to  
reduced UV irradiation and {\it in situ} mechanical heating,
both of which are driven by accretion
\citep[e.g.,][and references therein]{adamkovics2016}, 
potentially reducing the column density of the warm disk 
atmosphere.
The detectable CO emission of the low accretion rate CTTS, 
which have accretion rates similar to that of the TOs, implies 
that this scenario requires more than a reduced accretion rate 
to explain the reduction in CO emission. 

Empirically, the CO emission of the TOs in our sample does not appear to be 
much cooler than that of CTTS. 
Like the low accretion rate CTTS, the TOs LkCa~15 and GM~Aur have 
low-J and high-J lines of comparable equivalent width (Fig.~5). 
Only UX~Tau~A has stronger low-J lines than high-J lines
signaling an unusually low temperature.
\citet{salyk2009} previously noted that TOs typically do 
not show high-J lines, contrary to our results. The difference 
might be due to the range of $J$ sampled in our respective 
spectral settings. Our high-J lines are slightly lower in 
excitation than those studied by \citet{salyk2009}.

As a variant of this idea, shadowing effects might reduce the irradiation 
heating of the disk atmospheres of TOs and decrease their CO emission. 
Along these lines, \citet{espaillat2011} have noted a 
``see-saw'' behavior in the variability of pre-transition disks (a subset of TOs): 
when the near-infrared excess increases, the mid-infrared excess decreases. 
\citet{espaillat2011} can reproduce this behavior, observed for 
LkCa~15, UX~Tau~A, and IP~Tau, by 
changing the height of the inner disk edge near the dust 
sublimation radius (at $\sim 0.1$\,AU) 
by $\sim 22$\%, 17\%, and 17\% respectively.
When the height of the inner disk edge is larger, the emission
at the wavelengths where the inner disk edge dominates the emission 
($2-8\micron$) is higher and casts a larger shadow on the outer 
disk. Reduced irradiation heating of the shadowed region would 
reduce the column of warm CO-emitting gas.  
The changing height of the inner disk edge can be driven by varying 
accretion: a higher accretion rate raises the surface density of 
the inner disk, which increases both the IR excess and the height at which 
the inner edge of the disk becomes optically thick to stellar radiation. 

Some of the properties of the TOs in our sample are 
consistent with shadowing; i.e.,  
the narrow CO line profiles of the TOs locate the emission at larger radii
than the dust sublimation radius. 
For LkCa~15 and UX~Tau~A, the HWHM of their CO emission corresponds to 
emitting radii of 0.35\,AU and 2.5\,AU respectively given their 
stellar masses \citep[$1.05 M_\odot$ and $1.58 M_\odot$;][]{andrews2013}
and inclinations \citep[$i=49$ and $i=35$, respectively;][]{andrews2011b}, 
larger than the radius of the inner disk edge.  
However, the HWHM of the CO emission from GM~Aur also corresponds to 
a large emitting radius of 2\,AU given its 
stellar mass \citep[$1.35 M_\odot$;][]{andrews2013}
and inclination 
\citep[$i=56$;][]{andrews2011b} 
despite the lack of an optically thick inner disk edge that could 
possibly produce any shadowing. 
In addition, the variability seen in Figure 8 is inconsistent with 
the shadowing 
scenario. When $r_M$ increases significantly (which would correspond 
to the increasing height of the inner disk edge in this scenario), 
the CO emission is either unaffected (UX~Tau~A) or increases rather 
than declining as a result of greater shadowing (LkCa~15). 

To summarize, it seems plausible that 
the reduced CO emission from TOs 
results from chemical and/or dynamical effects. 
The low grain abundance of the inner disk of TOs is likely to 
reduce the abundance of CO close to the star, reducing the 
strength of the CO emission. 
Low mass planets located within 0.3~AU of the star may also 
open narrow gaps in the inner disk, reducing the CO emitting 
area.  
Detailed thermal-chemical models of disk atmospheres with low 
grain abundances and 
studies of the impact of low mass planets on inner disks 
are needed to further explore these interpretations.

\subsection{Residual Gas in WTTS Disks?}

To explore whether our results place any useful constraints
on the lifetime of gaseous inner disks,
we need to understand how our upper limits on CO emission strength 
translate into corresponding upper limits on gas mass.  
An important issue is how the CO abundance in the disk is affected by 
a diminishing grain abundance and whether the CO that remains in the disk is 
hot enough to emit detectable ro-vibrational emission. 

While CO could be dissociated more readily when the inner disk has
less dust to shield the gas from dissociating stellar UV irradiation,
processes such as molecular self-shielding and the reformation of CO
can help to maintain a significant reservoir of CO even when grains 
are greatly reduced in abundance \citep{bruderer2013,adamkovics2014,adamkovics2016}.
H$_2$ is an important precursor molecule for the formation of CO.
In the absence of grains, H$_2$ may form on PAHs \citep{bruderer2013} or
through gas phase processes \citep{glassgold2009}.
Therefore, CO might persist in the inner disk even in a disk with low 
continuum opacity.

Because multiple factors beyond gas column density affect the CO abundance 
and temperature (UV luminosity, grain abundance, PAH abundance), 
we would ideally rely on theoretical models of the thermal-chemical
properties of dissipating disks for insight. Unfortunately, these
models are in their infancy and general models for low mass
stars are not available. \citet{gorti2011} studied the specific
case of TW~Hya. \citet{bruderer2013} has primarily focused on Herbig stars. 

In the absence of general models for T~Tauri stars, 
we use the \citet{bruderer2013} models
to illustrate how our results could be used to constrain the residual 
gas in WTTS when appropriate models are available. 
On the one hand, the high FUV luminosity of the Herbig model (and 
resulting photodissociation) tends to underestimate the disk 
CO abundance. 
On the other hand, the high (ISM) PAH abundance that is assumed 
(a mass ratio of 5\% relative to dust) tends to increase the CO 
emission because of the role of PAHs in absorbing UV photons, 
heating the gas through the photoelectric effect, and 
as an additional pathway for H$_2$ formation.
While PAHs have been found to be present in the dust cavity of 
one Herbig system \citep[IRS 48;][]{geers2007}, 
their abundance in T Tauri disks is uncertain. 

In the \citet{bruderer2013} reference model for a transition disk, the inner region
of the disk is reduced in dust abundance (or column density) by a
factor of $10^{-5}$ relative to its original value. The original disk
has a gas mass column density of $1700\gpersqcm$ at 0.5 AU and
a dust mass column density of $17\gpersqcm$ at 0.5 AU. The value of the
decrement is consistent with observations of transition disks.
\citet{andrews2011b} reports decrements of $\sim 10^{-5} - 10^{-6}$
for transition disks at sub-millimeter wavelengths.

Despite the large FUV luminosity of the reference model ($0.7 L_\odot$), 
CO survives in the inner disk.  
When the gas column density is reduced by a factor of 100 relative
to its original value (to $\sim 17 \gpersqcm$ at 0.5 AU),
the gas reaches a temperature of $\sim 1000$\,K and
a density $\sim 10^{11}\percc$, and 
the disk is expected to produce CO emission \citep[Fig.~4 of][]{bruderer2013}.
This is consistent with our observations:
we do see CO emission from transition disks with \Mband\ excess.

\citet{bruderer2013} also presents results for systems with even
larger decrements in the dust abundance.
When there is no dusty inner disk
(in the model the dust is reduced by a factor of 
$\delta_{\rm dust} = 10^{-10}$),
the CO abundance remains high at the disk mid-plane 
($x_{\rm CO} = 10^{-4}$) for
a decrement in the gas disk of 
$\delta_{\rm gas} = 10^{-4}$ or a gas column
density of $0.17\gpersqcm$ at 0.5\,AU).
The CO in the mid-plane is hot ($\sim1000$\,K) because of the
lack of dust-gas cooling
and the mid-plane is dense 
$\sim 10^{12}\persqcm$ (from Fig.~4 of Bruderer [2013], scaling 
down to the mid-plane).
As a result, the CO column within 1\,AU ($10^{17}-10^{18}\persqcm$) 
is likely to produce CO ro-vibrational emission.

When the gaseous disk has an even lower column density, 
(i.e., $\delta_{\rm dust} = 10^{-10}$, 
$\delta_{\rm gas} < 10^{-4}$), 
there is
no substantial CO reservoir in the inner disk
\citep[$N_{\rm CO} < 10^{14} \percc$; see also Fig 8 of][]{bruderer2013}.
Using these models as a guide, we would not expect to see CO
ro-vibrational emission for dust-free disks with 
gas column densities $\lesssim 0.17\gpersqcm$ at 0.5\,AU.
A better estimate for the limiting gas column for
detectable CO ro-vibrational emission could be made with models
that are more closely tailored to the properties of T~Tauri disks 
and which explore the sensitivity of the results to  
PAH abundance.

If our WTTS observations truly indicate a gas column density limit of
$< 0.17 \gpersqcm$ at 0.5\,AU, they help to constrain
the impact of residual gas on the eccentricities of forming planets.
Gravitational interactions between planets and the gas
damp eccentricity and inclination \citep{kominami2002,agnor2002}.
As a result, 
in closely packed planetary systems, a residual gas disk may 
play a role in suppressing planet-planet interactions and scattering
\citep{alexander2014}.
\citet{kominami2002} suggested that residual gas in the terrestrial
planet region at the level of $10^{-3} - 10^{-4}$ of the MMSN
could circularize the orbits of terrestrial planets on timescales
of $\sim 10$\,Myr \citep[in the absence of secular perturbations by
giant planets;][]{kominami2004},
potentially producing the low eccentricities observed in the
solar system. 
If the Solar System planets formed in a disk like the WTTS systems 
studied here, gas drag would be unable to circularize the planets in 
$\sim 10$\,Myr 
because little gas remains in the disk
($<10^{-4}$ of the MMSN) at ages $\lesssim 10$\,Myr.
More dilute reservoirs of gas ($<10^{-4}$ of MMSN), 
beyond our ability to probe, could reduce exoplanet eccentricities 
on longer timescales. 

The lack of evidence for residual gas disks surrounding the
WTTS, as traced by CO in our sample
(age 1--2\,Myr for Taurus and $\sim 10$\,Myr for TWA) does
not help to explain the slow rotation rates of older stars
\citep[see][for a review]{bouvier2014}.
\citet{gallet2013} have shown that the rotational periods
of stars as a function of age can be explained if solar-like stars
remain magnetically locked to their gaseous inner disks for
5 Myr on average. The rotation rates of the slowest 25\% of
sources in older clusters ($> 10$\, Myr) requires a longer disk locking
time of $\sim 7$\,Myr.
There is little evidence that gaseous disks survive
that long.
Very few young stars are accreting at 5 Myr of age.
Using H$\alpha$ emission as a diagnostic, 
\citet{fedele2010} found that 
95\% of solar-type stars do not show detectable accretion 
at 5 Myr.
The inference of negligible gaseous reservoirs in the inner disk
could be avoided if stellar accretion rates
are a poor tracer of inner disk gas, i.e., if a substantial
fraction of stars with negligible stellar accretion maintain a
gas-rich inner disk. However, we find no evidence for residual
gas disks surrounding (non-accreting) WTTS.
This descriptive result could be made more quantitative 
when there are theoretical predictions for the amount of residual gas 
needed for rotational breaking to occur. 

In contrast, if our observations indicate a gas column density limit of
$< 0.17 \gpersqcm$ at 0.5\,AU, they are not sensitive enough 
to determine whether residual gas can whisk away the dust debris 
generated by terrestrial planet formation and invalidate the 
commonly held assumption that warm debris disks are tracers of 
terrestrial planet formation.  
As discussed by \citet{kenyon2016}, 
if residual gas is present in the terrestrial planet region 
at a level $\gtrsim 10^{-5}$ of MMSN, the action of 
aerodynamic drag and radiation pressure can 
remove (or prevent the formation of) the debris particles  
that produce an observable excess. 

Measurements that are sensitive to residual gas at this level could 
weigh in on whether residual gas can resolve a new puzzle: 
the surprising discrepancy 
between the high frequency of Earth-mass exoplanets surrounding 
mature sun-like stars and the lack of evidence for the 
warm debris that their formation is expected to produce.  
The occurrence rate of Earth-mass exoplanets at orbital distances of 
0.4-1\,AU ($\sim 20$\%) is found to be much larger than the incidence 
rate of warm debris around solar-type stars at ages 
$\lesssim 10$\,Myr (few \%). As discussed by \citet{kenyon2016}, 
a simple way to resolve this discrepancy is if a dilute reservoir 
of gas ($10^{-5}$ of MMSN) persists for $\sim 10$\,Myr 
and removes the debris produced by planet formation. 
More sensitive measures of residual gas than those obtained here, 
as well as thermal-chemical models tuned to the conditions of 
dissipating T Tauri disks, 
are needed to test this hypothesis. 
Alternative explanations include the possibility that 
rocky Earth-mass planets are not as common as we think, 
or that terrestrial planets form very differently than currently 
imagined, through a process that leaves behind much less debris.

Our results complement stellar accretion rates as a
diagnostic of the evolution of gaseous disks.
While stellar accretion rates measure the gas mass flux
reaching the star, if accretion is halted for some reason,
residual gas could remain in the disk despite a low stellar
accretion rate.
Our results suggest that this outcome is not common: 
we find that CO ro-vibrational emission is detected only
from systems with detectable stellar accretion. 
However, we cannot rule out the presence of small 
column densities of residual gas that, while limited in 
mass, may nevertheless have a significant impact on our 
ability to detect terrestrial planet formation in action.

Complementary results have been obtained in studies that use
alternative {\it in situ} diagnostics to track the 
evolution of gaseous disks \citep[see][for a recent review]{alexander2014}.
Interpreting {\it Spitzer Space Telescope} upper limits on
mid-infrared H$_2$ line fluxes, \citet{pascucci2006} inferred
gas column density limits $< 10^{-4}$ of MMSN at disk 
radii $>1$\,AU 
in the sample of older stars ($>$ few Myr) studied by FEPS.
In contrast to the CO fundamental emission studied here, 
the FEPS diagnostics are insensitive to, and potentially 
compromised by, gas close to the star. In the \citet{gorti2004} 
models used to interpret the FEPS upper limits, 
if the gaseous disk extends in to small radii, 
the irradiation heating is deposited primarily close to the star 
and most of the disk is too cool to emit in the {\it Spitzer} lines.
Thus our results complement the FEPS results in probing 
distances similar to the orbital radii of known exoplanets 
and the radii where star-disk locking potentially occurs. 

UV fluorescent H$_2$ emission, which is excited by Ly$\alpha$
and arises from similar disk radii as CO fundamental emission, 
is detected from CTTS, but not WTTS
\citep{ingleby2009,france2012}.
One difficulty with interpreting the lack of fluorescent
H$_2$ emission as evidence for the lack of residual gas is
that the fluorescent emission requires both Ly$\alpha$ irradiation 
and hot H$_2$ (2000K), neither of which may be abundant in
(non-accreting) WTTS.

An alternative UV diagnostic was pursued by
\citet{ingleby2012}, who used the 1600\AA\
feature to infer a short gas dissipation timescale for WTTS disks.
The 1600\AA\ feature is attributed to electron impact excitation
of H$_2$ \citep{bergin2004,ingleby2009,france2011}.
Because the hot electrons invoked are thought to be produced by
X-ray irradiation, and because stellar X-ray irradiation is
long-lived (on timescales of tens of Myr), the 1600\AA\ feature
is thought to potentially probe residual gas in a non-accreting disk.
To confirm this picture, it would be useful to develop models
that show that the presence of a residual gas disk
would produced a detectable 1600\AA\ feature flux
given the typical X-ray luminosity of WTTS.


\section{Summary and Conclusions}

We have carried out a sensitive search for residual gas and dust 
in the terrestrial planet region surrounding young stars 
ranging from a few Myr (Taurus) to $\sim 10$\,Myr (TWA) in age. 
Using high resolution $4.7\micron$ spectra of transition objects 
and weak T~Tauri stars, 
we searched for weak continuum excesses and CO fundamental emission 
after making a careful correction for the stellar contribution 
to the observed spectrum. 

We find that the CO emission from transition objects is weaker 
and located further from the star than CO emission from 
non-transition T~Tauri stars with similar stellar accretion rates.  
The difference is possibly a chemical effect, i.e., 
transition objects have robust gaseous disks close to the star, 
but the CO abundance there is low and weak CO emission results. 
Alternatively, the gaseous disk close to the star may be truncated 
by the presence of close-in companions. 

Some weak T Tauri stars are found to have weak \Mband\ veiling, but none show 
CO fundamental emission down to low flux levels 
($5\times 10^{-20}-10^{-18}\, {\rm W\,m^{-2}}$).  
These limits 
can potentially lend insights into 
the ability of residual gas disks to 
spin down stars,   
circularize the orbits of terrestrial planets, 
and whisk away the dusty debris that is expected to serve as a 
signpost of terrestrial planet formation. 
To impose these limits, we need to convert our upper limits on 
CO flux into gas column density upper limits. This conversion 
requires detailed models of the CO abundance and 
gas temperature in disk atmospheres with low grain abundance, 
which are not yet generally available. 

If the models of \citet{bruderer2013}, which focus on Herbig stars, 
are a guide, our limits on CO flux translate into a gas 
column density of $\lesssim 10^{-4}$ of MMSN at 0.5\,AU. 
Gas disks this tenuous are unlikely to play a major role 
in circularizing terrestrial planets. 
At a gas column density of $\sim 10^{-4}$ of MMSN, 
circularizing terrestrial planets through gas drag 
requires timescales longer than 10 Myr. 
The lack of evidence for residual gas in WTTS, as traced by CO also does not help to 
explain the slow rotation rates of older stars \citep{bouvier2014}, 
which require solar-mass stars to remain locked to their 
disks for $\sim 5$\,Myr \citep{gallet2013}. 
This constraint can be made more quantitative when there are 
theoretical predictions for the amount of residual gas that is 
needed for rotational breaking to occur. 

However, the same upper limit of $\lesssim 10^{-4}$ of MMSN 
is not sensitive enough to 
weigh in on whether residual gas whisks away 
the dusty debris signature of terrestrial planet formation. 
As described by \citet{kenyon2016}, 
the unexpected discrepancy between the high frequency of 
Earth-mass exoplanets within an AU of mature sun-like 
stars ($\sim 20$\%) and low incidence rate (few \%) 
of the warm debris that their formation is expected to produce 
could be explained as a consequence of 
a long-lived dilute reservoir of residual gas 
(at a level of $\gtrsim 10^{-5}$ of MMSN) 
that removes the debris through aerodynamic drag and radiation pressure. 
This explanation is perhaps more palatable than 
alternative explanations which include the 
possibility that rocky Earth-mass planets are not as common as 
currently estimated or that terrestrial planets form via a much 
different (neater) process than is currently imagined, producing much less debris. 

To pursue these topics more quantitatively,   
we need to develop thermal-chemical models for T~Tauri disks 
that explore their structure at very low grain abundance. 
It is also of interest to improve our sensitivity to weak emission 
from CO and other tracers of gas in the terrestrial planet region. 
Both higher signal-to-noise spectra and more highly optimized 
stellar atmosphere corrections would be valuable in this regard.

\acknowledgments

The authors wish to recognize and acknowledge the very significant
cultural role and reverence that the summit of Mauna Kea has always
had within the indigenous Hawaiian community.  We are most fortunate
to have the opportunity to conduct observations from this mountain.
We thank the Keck Observatory staff who provided support and
assistance during our NIRSPEC run.  We thank the anonymous referee for a thorough 
reading of the manuscript, whose comments helped improve the paper.
Financial support for GWD was provided in part by the 
NASA Origins of Solar Systems program NNH10A0061.  
JSC acknowledges 6.1 funding for basic research in infrared astronomy 
at the Naval Research Laboratory.
JN acknowledges the stimulating research environment 
supported by NASA Agreement No.\ NNX15AD94G to the 
``Earths in Other Solar Systems'' program. 

{\it Facility:} \facility{Keck:II(NIRSPEC)}.

\clearpage


\clearpage




\begin{figure}
\figurenum{1}
\epsscale{1.0}
\plotone{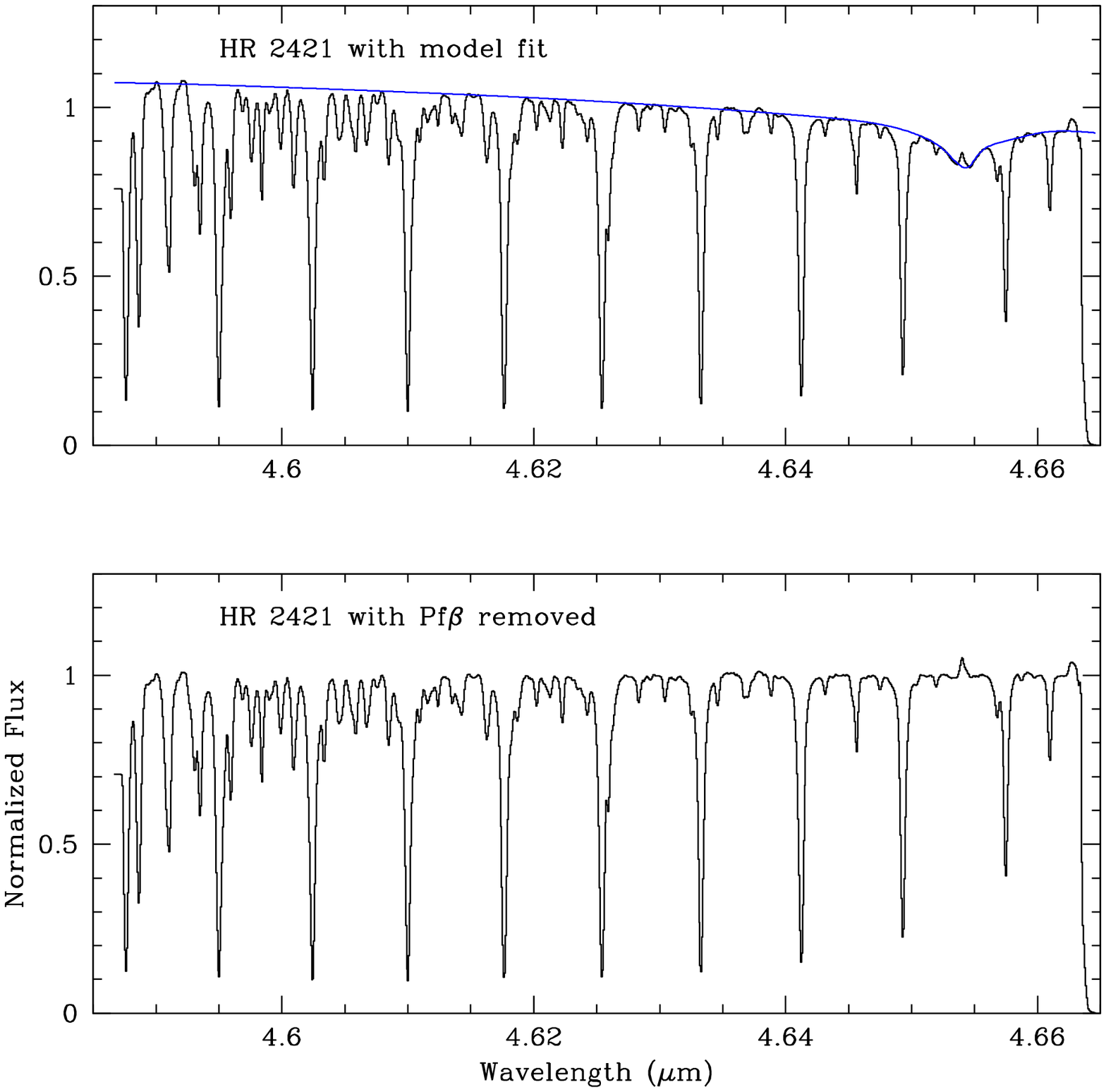}
\caption[]
{
The observed spectrum of an early type standard star, HR~2421 (A0 IV) showing broad Pf$\beta$
absorption (black plot in top panel).  A spline is fit to the $4.6\micron$ continuum of HR~2421 to model the broad shape of the  Pf$\beta$ absorption line (blue line in top panel). Dividing the observed spectrum by the spline fit removes the broad Pf$\beta$
emission in the observed telluric spectrum (bottom panel), which is then used to remove the narrow telluric features  without altering the shape of the  Pf$\beta$ emission that may be present in our YSO targets.
}
\end{figure}

\begin{figure}
\figurenum{2}
\epsscale{1.0}
\plotone{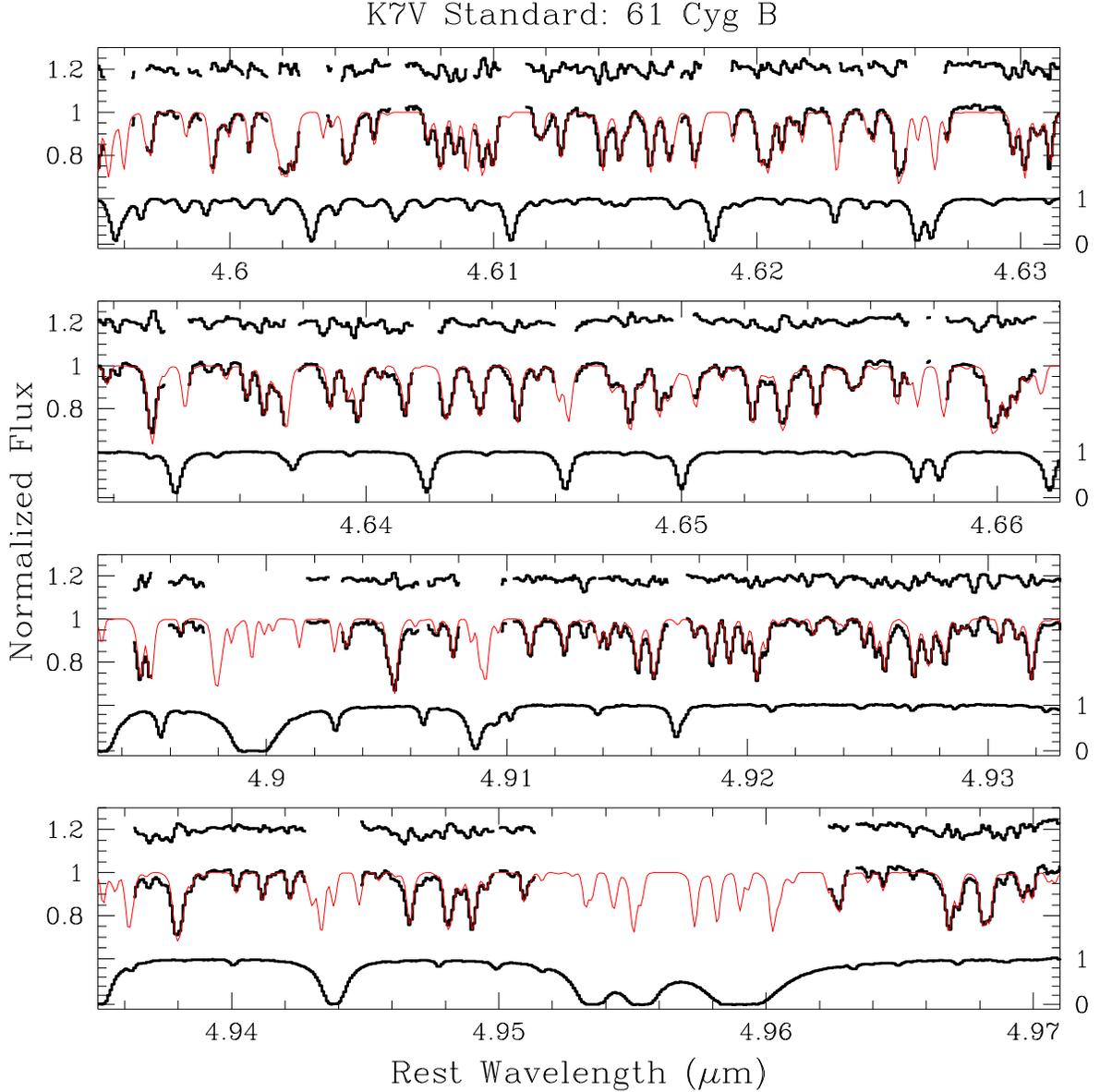}
\caption[]
{
Comparison of model fit and observed spectrum of 61~Cyg~B, 
a K7V standard star.  
In each panel, the observed spectrum (thick black histogram)
and model fit (thin red line) are shown, along with the residuals 
(upper black histogram) and the telluric spectrum (lower black line) 
for comparison.  The 
observations are well fit by a model with $\teff$ = 4000K, 
$\logg$ = 4.5 $\gpersqcm$, [Fe/H]$=-0.27$, and no continuum veiling, 
consistent with the properties of a
late-type main-sequence dwarf.  We searched a range of veiling
values ($\mveil = -0.3$~to 0.4 in steps of 0.1), and found the best
fit at zero veiling.  Many of the observed residual absorption
features not fit by the model in the lower two panels 
(i.e., 4.91--4.94\,$\micron$,  
4.945--4.95\,$\micron$, and 4.963--4.97\,$\micron$) 
are consistent
with water lines identified in the solar sunspot atlas.
}
\end{figure}

\begin{figure}
\figurenum{3a}
\epsscale{1}
\plotone{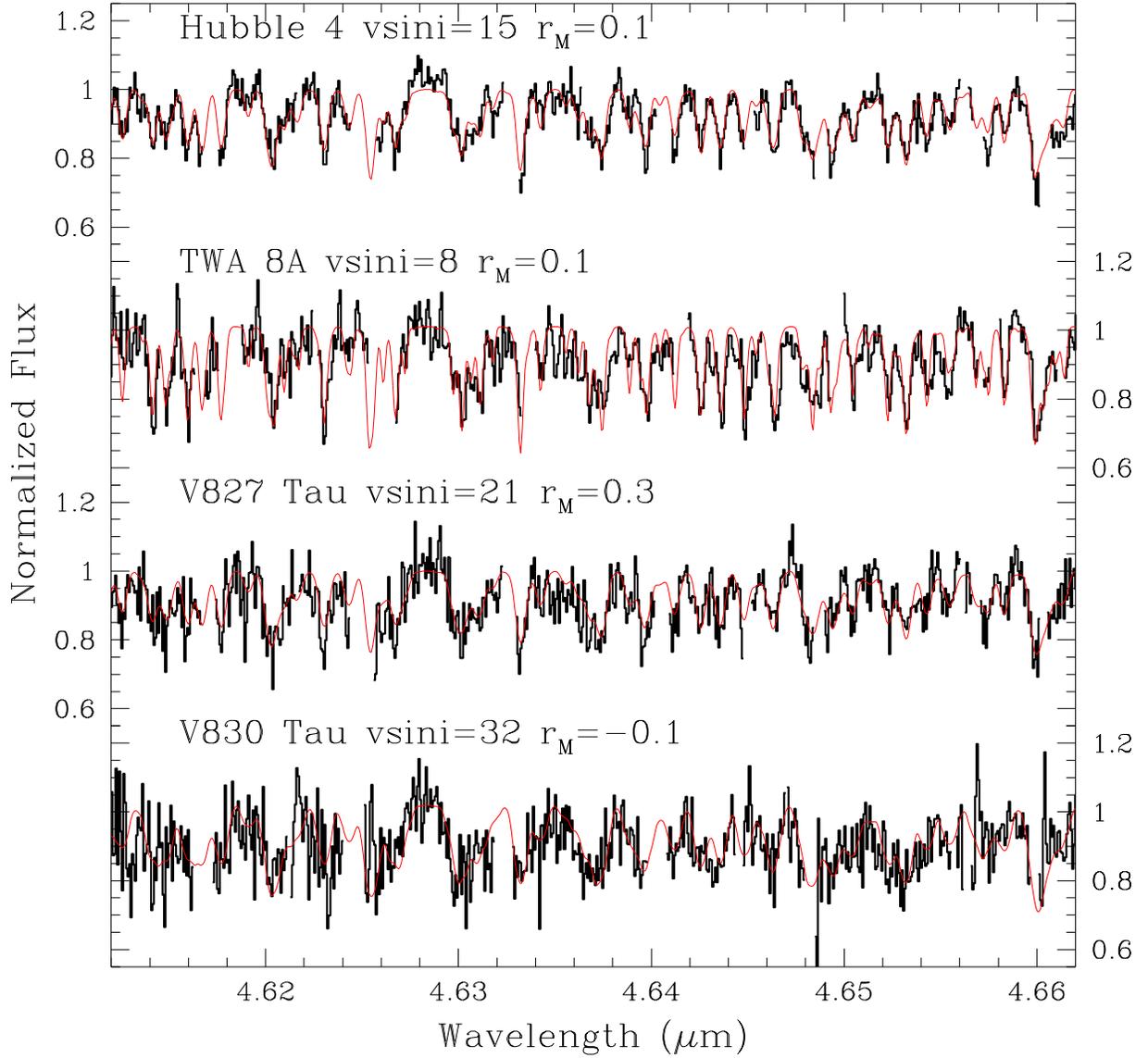}
\caption[]
{
Model fits (thin red lines) to the $4.6\micron$
spectra of representative sources (thick black histogram) 
that have little or no veiling ($\mveil \le 0.3$).  
The model fits to these WTTS do a reasonable job fitting 
a range of $\vsini$ ($8-32~\kms$, see Tables 1 and 2) 
at both $4.6\micron$ and $4.9\micron$. 
Spectral regions with poor telluric transmission ($\le 80\%$) 
have been excised from the plot. }
\end{figure}

\begin{figure}
\figurenum{3b}
\epsscale{1.0}
\plotone{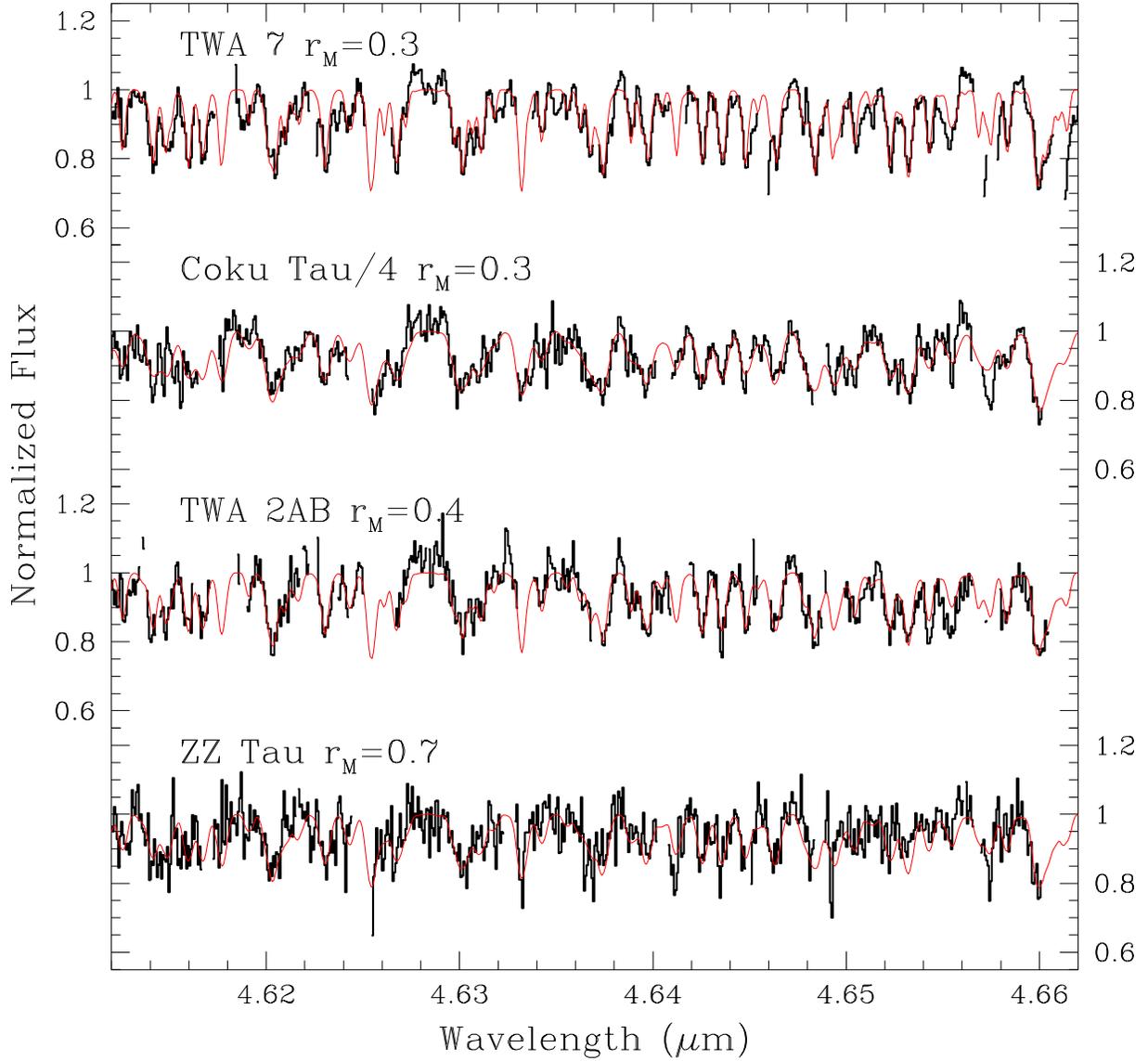}
\caption[]
{
As in Fig. 3a but for sources with
veiling $\mveil$ from 0.3--0.7, including Coku~Tau/4, a transition
object, and ZZ~Tau, a CTTS.}
\end{figure}

\begin{figure}
\figurenum{3c}
\epsscale{1.0}
\plotone{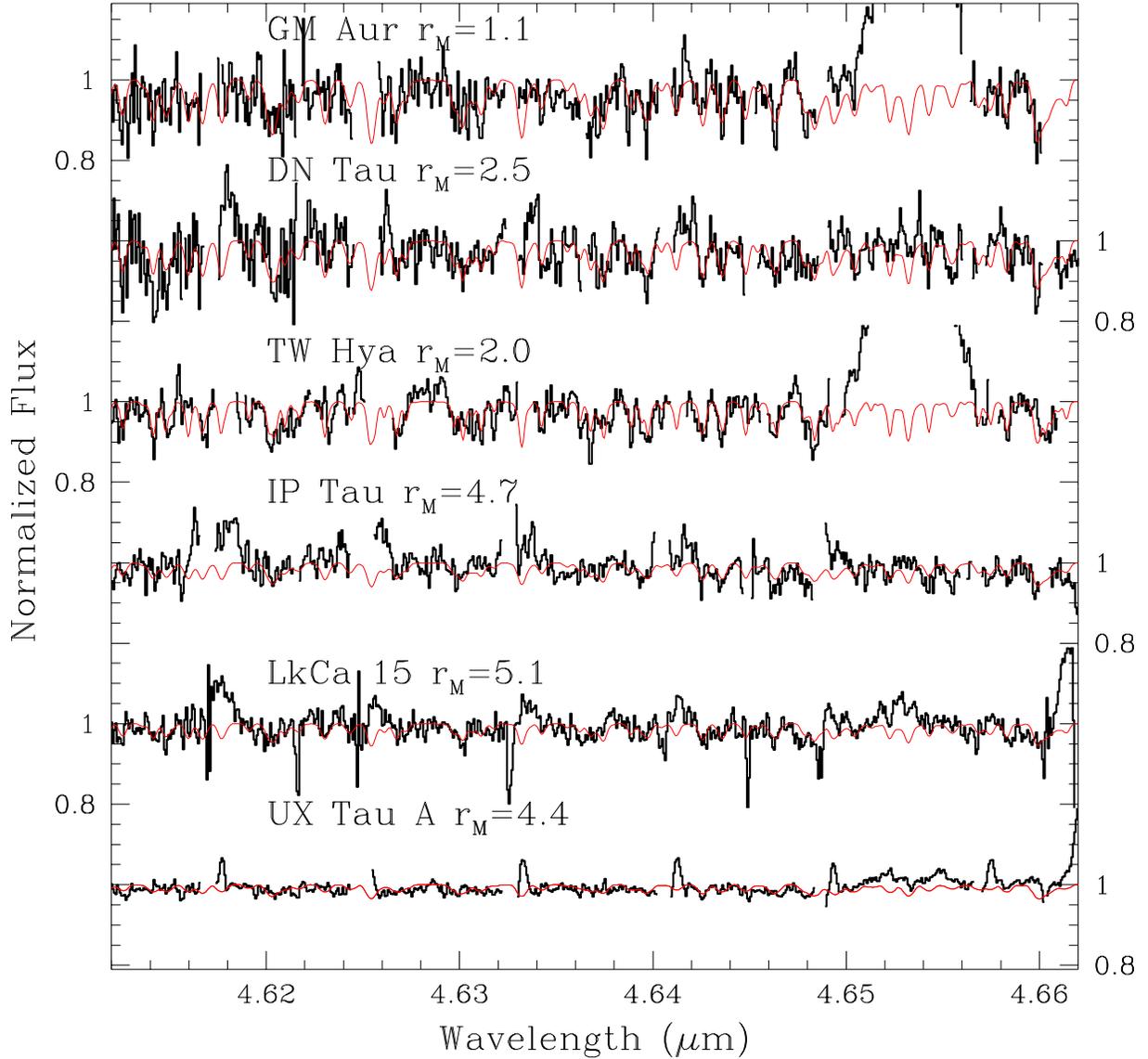}
\caption[]
{
As in Fig 3a but for sources with high veiling
($\mveil = 1.1-5.1$).  Numerous absorption lines in the model
fit are coincident with detected emission lines in the observed
spectra, underscoring the utility of accurately modeling the stellar
component in order to detect weak emission that originates from the
warm inner disk.  Strong Pf$\beta$ emission is also seen at
$4.652\micron$ in the spectra of TW~Hya and GM~Aur. } \end{figure}

\begin{figure}
\figurenum{4a}
\epsscale{1.0}
\plotone{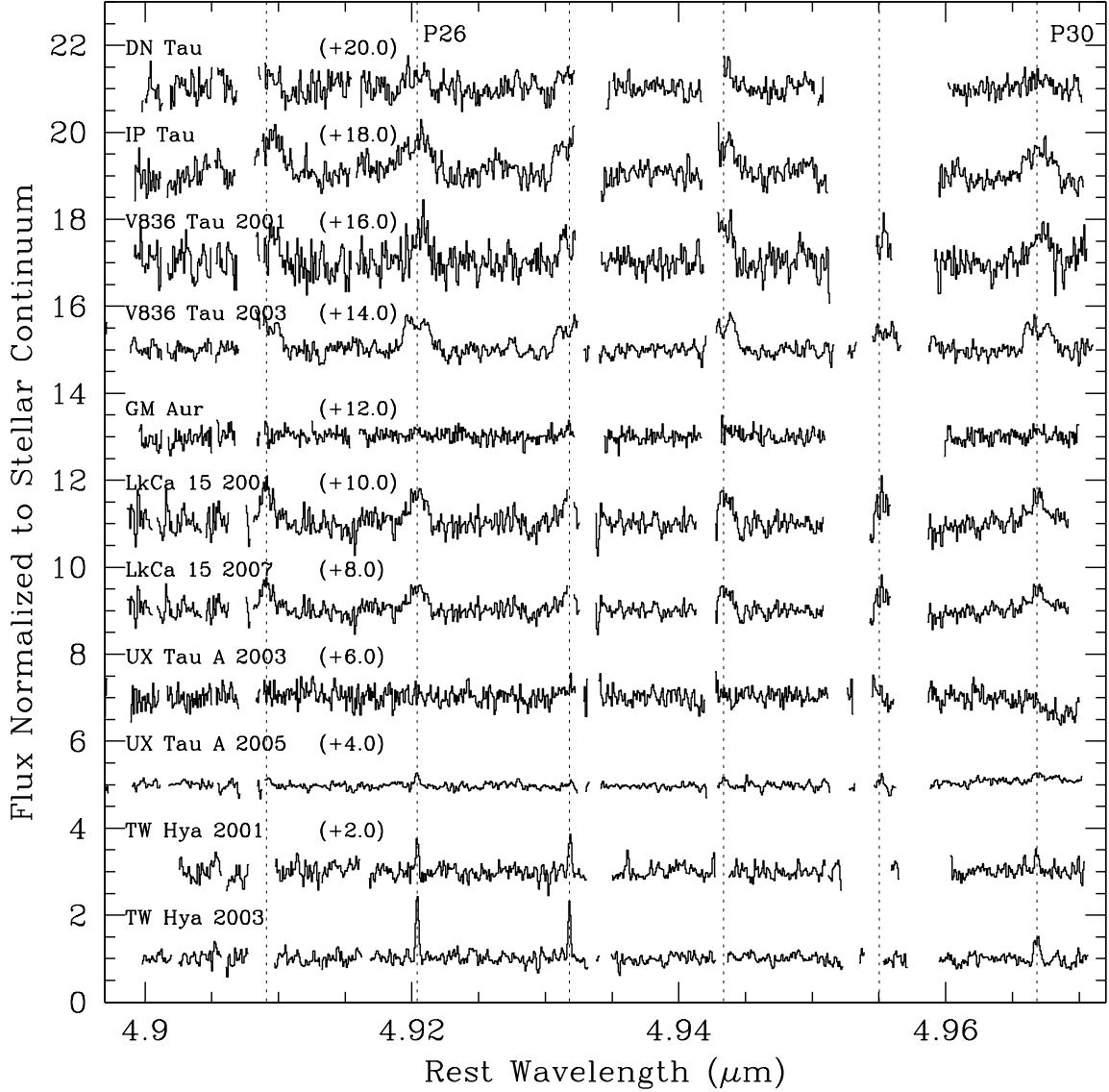}
\caption[]
{Observed \Mband\ spectra with the model veiled stellar photosphere
subtracted and replaced with a featureless continuum of equal strength.  
An arbitrary continuum offset  has been added to each spectrum (indicated in parentheses). 
Only sources with detected high-J (P25-P30) CO emission are shown.
The residual spectra of CTTS and TOs show CO fundamental emission.  
The high-J lines (P25--P30) shown reveal a range of line strengths 
and widths.  Spectral sub-regions with telluric transmission less than
80\% have been excised from the plot. 
}
\end{figure}

\begin{figure}
\figurenum{4b}
\epsscale{1.0}
\plotone{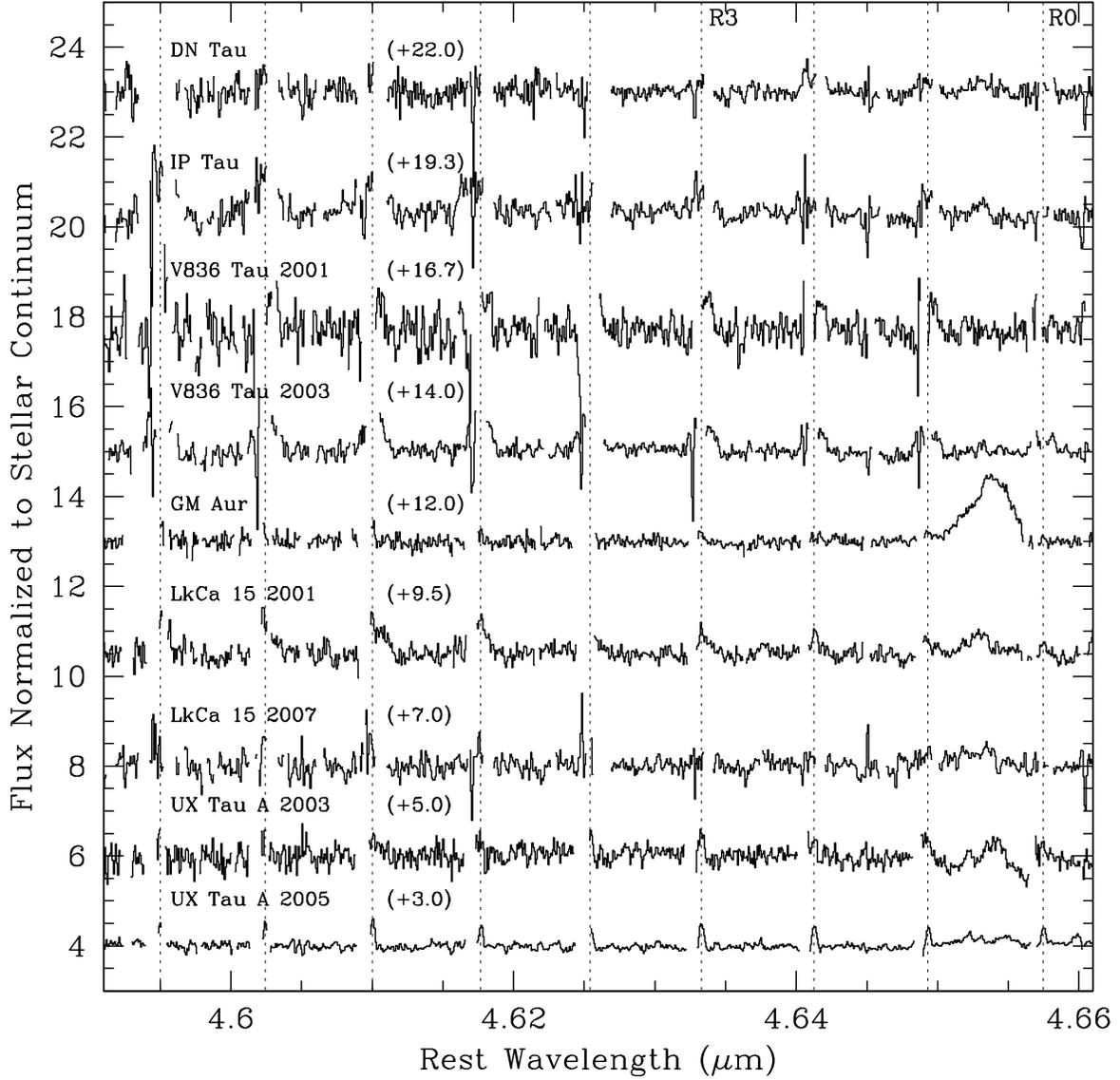}
\caption[]
{As in Fig.~4b but for low-J lines (R8--R0). 
Pf$\beta$ emission is evident in some sources at
$4.653\micron$.  
TW~Hya, which shows high-J CO emission in our data (Figures 4a and 5), 
is not included here; 
the source was not adequately shifted in velocity out of the 
telluric CO, resulting in poor telluric correction of the low-J CO spectrum.
} \end{figure}

\begin{figure}
\figurenum{5}
\epsscale{1.0}
\plotone{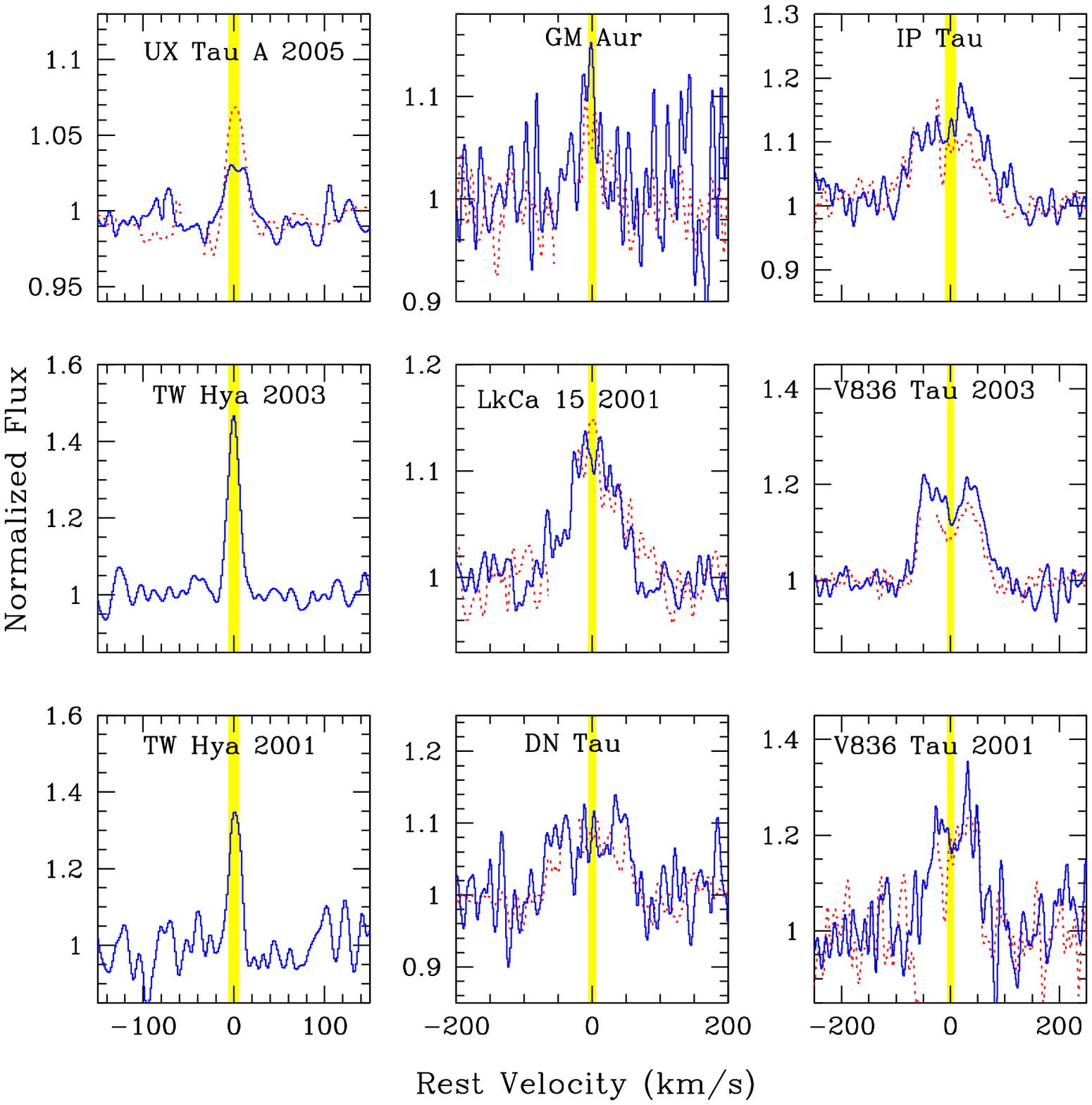}
\caption[]
{
CO line profiles, obtained by averaging low-J R-branch lines (dotted
red histogram) and high-J P-branch lines (solid blue histogram).  
FWHM line profiles and equivalent widths are measured
from these averaged profiles (see Table 3).  
The vertical line has the width of a spectral resolution 
element. Note that the velocity scale differs in each column. 
} \end{figure}

\clearpage
\oddsidemargin=-1cm
\tabletypesize{\scriptsize}

\begin{figure}
\figurenum{6}
\epsscale{1.0}
\plotone{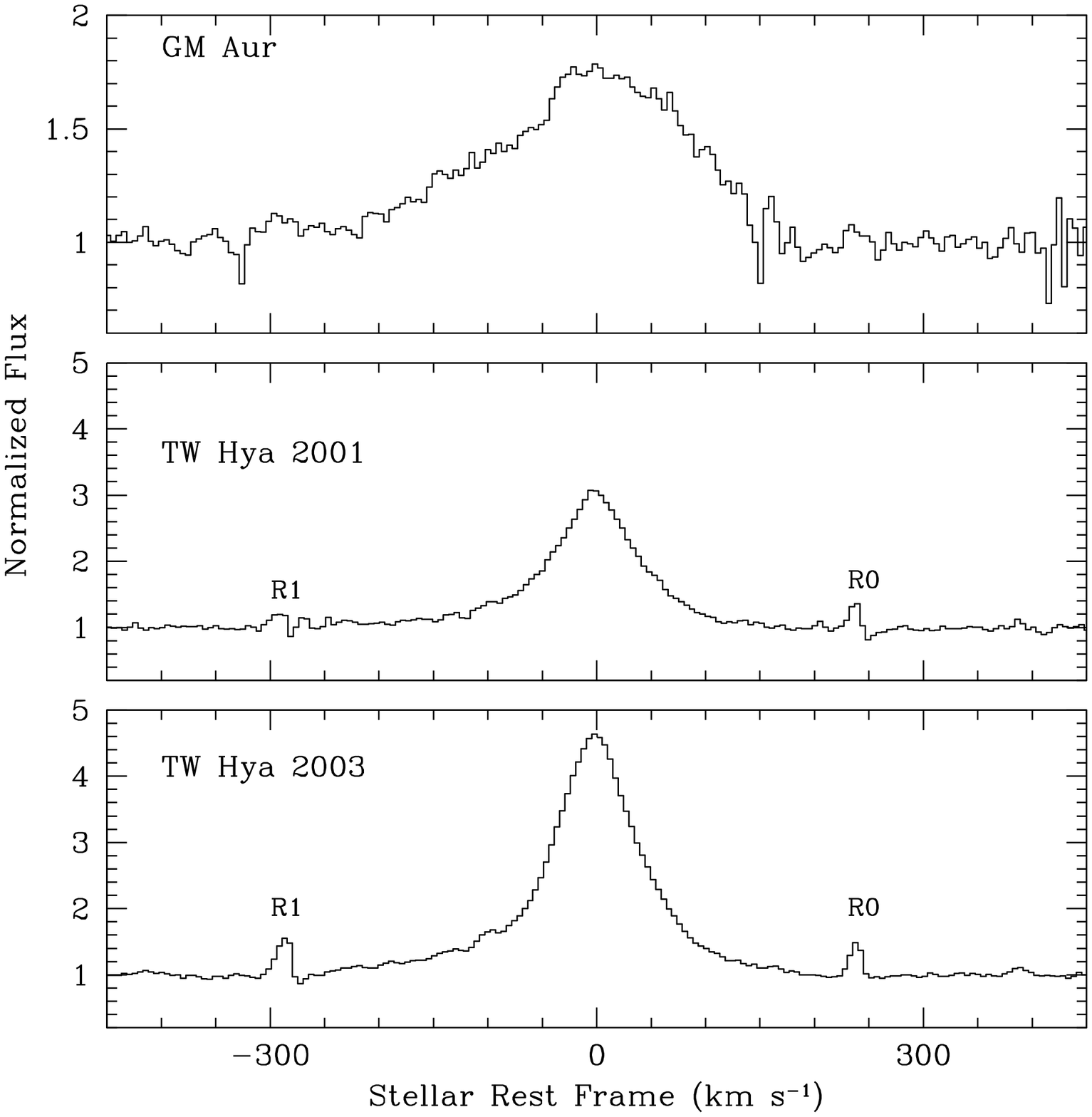}
\caption[]
{
Strong Pf$\beta$ line emission observed in our study.  
R0 and R1 CO emission lines are indicated in the TW~Hya spectrum.   
%
%
} \end{figure}

\begin{figure}
\figurenum{7}
\epsscale{1.0}
\plotone{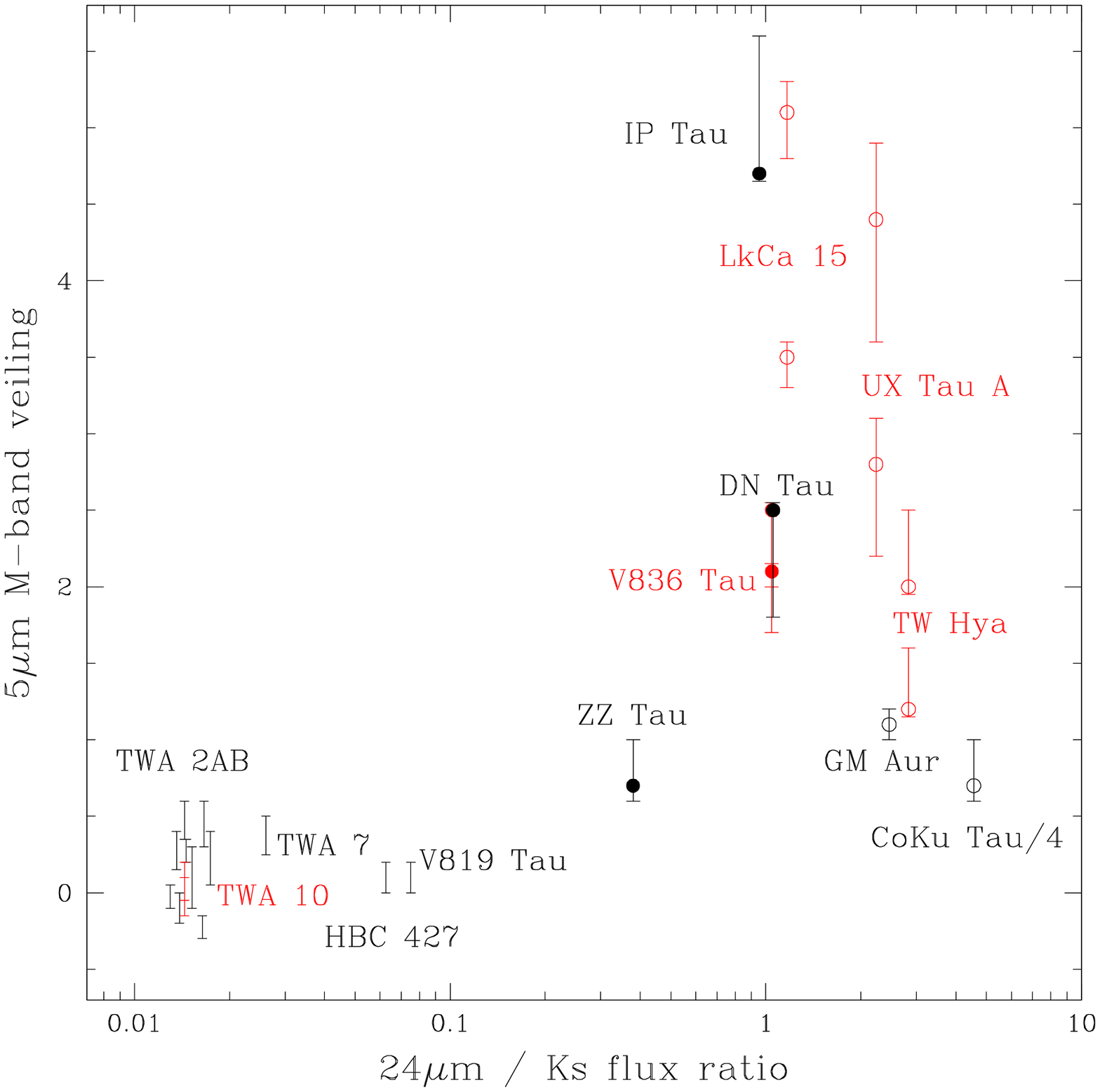}
\caption[]
{
Measured \Mband\ veiling vs.\ mid-infrared color excess.  The
$24\micron$ fluxes \citep{low2005,rebull2010,furlan2006,furlan2011}
have been normalized by the 2MASS $K_s$ flux value.   Red
points indicate observations made in more than one epoch. Low
accretion rate CTTS are shown as solid symbols, while TOs are displayed as open
symbols. The tight cluster of objects in the lower left corner are TWA and Taurus
WTTS sources in our sample which lack
significant \Mband\ veiling.  The asymmetric error bars stem from
unequal changes in the depths of stellar lines for $\teff$ and
$\logg$ pairs that span the uncertainties in those quantities.
} \end{figure}

\begin{figure}
\figurenum{8}
\epsscale{1.0}
\plotone{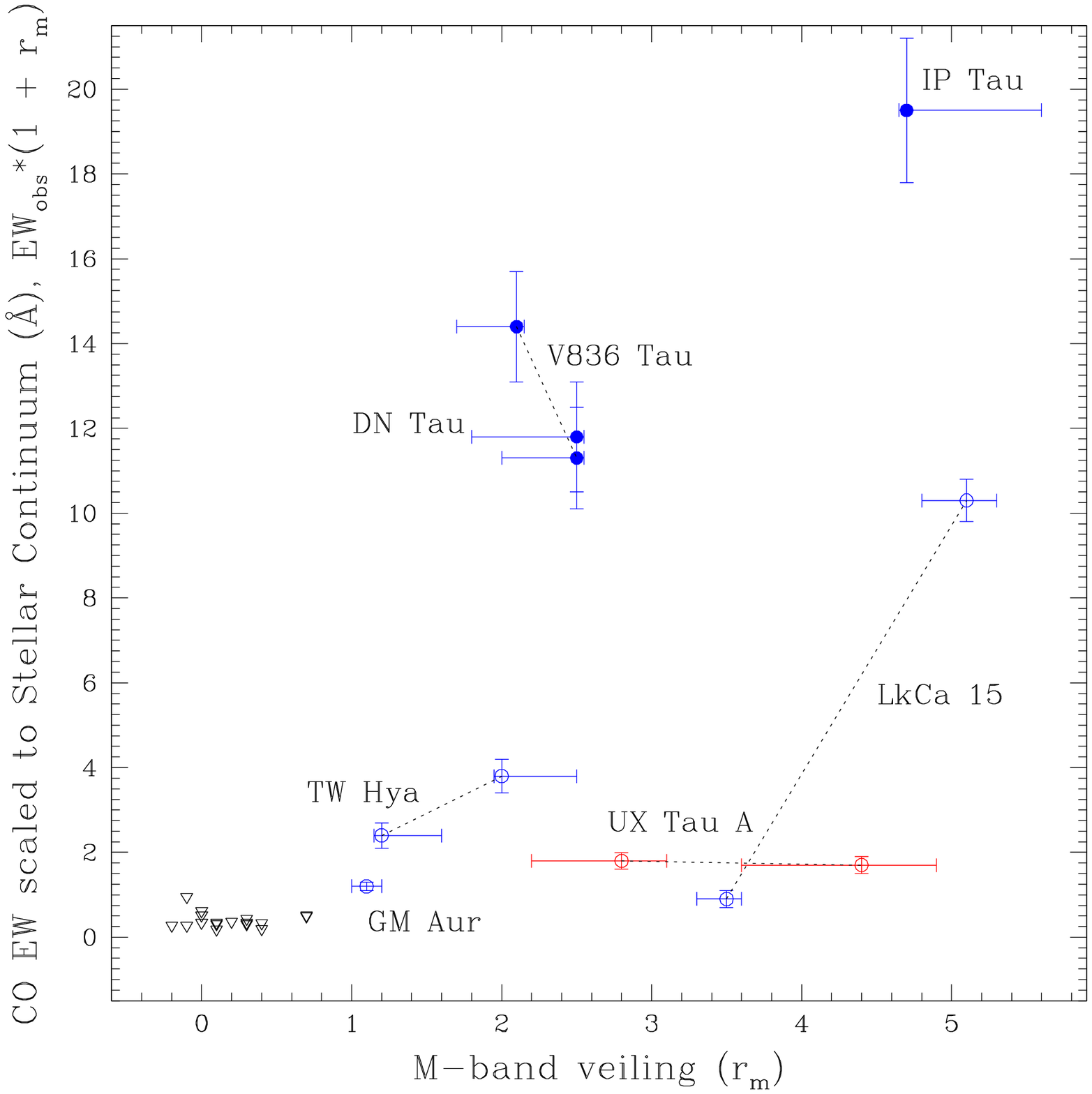}
\caption[]
{
Measured \Mband\ veiling vs. CO line flux relative to the stellar continuum 
for CTTS (filled circles), TOs (open circles), and WTTS (inverted black triangles).
CO line strengths are measured from combined R-branch (red) or
P-branch (blue) points, respectively.  
Objects with more than one observational
epoch are connected by a dashed line.  
WTTS, which had no CO emission and little or no veiling, are 
placed at their 1$\sigma$ emission detection
upper limit for CO equivalent width (lower left corner of plot).
} \end{figure}

\begin{figure}
\figurenum{9}
\epsscale{1.0}
\plotone{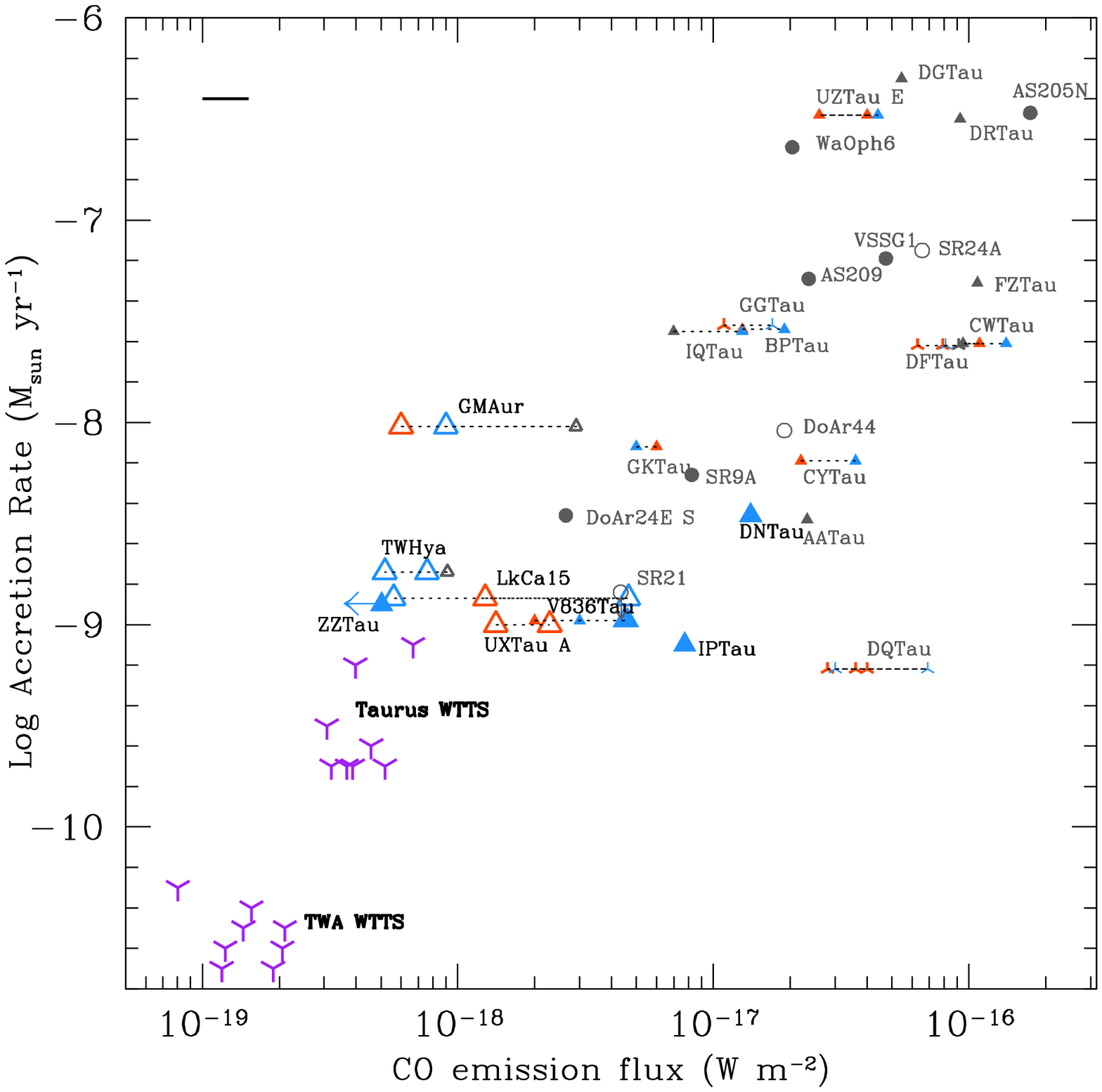}
\caption[]
{
CO emission flux vs.\ stellar accretion rate for 
TW~Hya and sources in Taurus (triangles) and Ophiuchus (circles).  
Low-J (R0--R5, orange), moderate-J (P8, gray),
and high-J lines (P25--P29, blue) are shown for 
CTTS (solid symbols), TOs (open symbols), and 
binary sources (skeletal triangles).  
Objects with multi-epoch observations are connected by a horizontal dotted line.
The emission flux for TWA and  Ophiuchus sources have been scaled
to the same distance as Taurus for comparison. Larger symbol sizes
denote objects from this study. A representative 20\% error
bar in the flux calibration is shown as a horizontal bar.
Upper limits for stellar accretion rates and 
CO line fluxes (an average of the low-J and high-J values) in 
TWA and Taurus are shown for comparison (inverted purple skeletal 
triangles).
Accretion rate and flux values for
targets from other studies are taken from
\citet{espaillat2010,eisner2005,valenti1993,natta2006,
brown2013,najita2007}.
}
\end{figure}

\clearpage



\begin{deluxetable}{lclcccccccccc}
\tablecolumns{13}
\tablewidth{0 pt}
\tablecaption{Journal of Observations and Stellar Parameters}
\tablehead{
\colhead{Object}	&
\colhead{Type$\tablenotemark{a}$}	&
\colhead{Obs Date}	&
\colhead{Int. Time}	&
\colhead{SNR}	&
\colhead{Slit Width}	&
\colhead{SpT} &
\colhead{$\teff$}	&
\colhead{Ref}	&
\colhead{$\logg$}	&
\colhead{Ref}	&
\colhead{$\vsini$}	& 
\colhead{Ref} \\
\colhead{} &
\colhead{} &
\colhead{(UT)} &
\colhead{(min.)}	&
\colhead{4.630 $\micron$}	&
\colhead{(arcsec)}	&
\colhead{} &
\colhead{(K)}	&
\colhead{}	&
\colhead{($\gpersqcm$)}		&
\colhead{}	&
\colhead{($\kms$)}			& 
\colhead{} \\
}

\startdata

DN Tau		& C  & 2007 Jan 2+3		& 64   & 22 & 0.43 & M0.3 & 3846 & 1 & 3.6 & 1 & 12 & 10 \\
IP Tau		& C  & 2003 Jan 13		& 48   & 35 & 0.72 & M0.6 & 3792 & 1 & 3.9 & 1 & 12 & 10 \\
V836 Tau		& C  & 2001 Jan 20		& 16   & 17 & 0.43 & M0.8 & 3756 & 1 & 4.0 & 1 & 13 & 10 \\
V836 Tau		& C  & 2003 Jan 11		& 40   & 30 & 0.43 & M0.8 & 3756 & 1 & 4.0 & 1 & 13 & 10 \\
ZZ Tau		& C  & 2001 Jan 20		& 12   & 20 & 0.43 & M4.3 & 3127 & 1 & 3.0 & 1 & 22 & 11\\
\\
CoKu Tau/4	& W,T  & 2005 Feb 18+19	& 190  & 30 & 0.58 & M1.1 & 3704 & 1 & 3.9 & 1 & 26 & 10 \\
GM Aur		& C,T  & 2001 Jan 19	& 16   & 25 & 0.43 & K6.0 & 4115  & 1 & 4.1 & 1 & 15 & 10 \\
LkCa 15		& C,T  & 2001 Jan 19	& 16   & 50 & 0.43 & K5.5 & 4163  & 1 & 3.9 & 1 & 14 & 10 \\
LkCa 15		& C,T  & 2007 Jan 2+3	& 84   & 40 & 0.43 & K5.5 & 4163  & 1 & 3.9 & 1 & 14 &  10 \\
UX Tau A		& C,T  & 2003 Jan 11	& 48   & 30 & 0.43 & K0 & 5250$\tablenotemark{b}$ & 2 & 4.5$\tablenotemark{b}$ & 2 & 24 & 10 \\
UX Tau A		& C,T  & 2005 Feb 19	& 56   & 90 & 0.43 & K0 & 5250$\tablenotemark{b}$ & 2 & 4.5$\tablenotemark{b}$ & 2 & 24 & 10 \\
\\
HBC 427		& W  & 2003 Jan 13		& 16   & 15 & 0.58 & K6.0 & 4115 & 1 & 3.8 & 1 & 10 & 10 \\
Hubble 4		& W  & 2003 Jan 13		& 24   & 25 & 0.58 & K7 & 4158$\tablenotemark{b}$	 & 3 & 3.6$\tablenotemark{b}$ & 3 & 15 & 3 \\
IW Tau		& W  & 2003 Jan 12		& 44   & 15 & 0.72 & M0.9 & 3738 & 1 & 3.7 & 1 & 9  & 10 \\
LkCa 19		& W  & 2003 Jan 12		& 48   & 15 & 0.43 & K0 & 5100$\tablenotemark{b}$ & 2 & 3.8$\tablenotemark{b}$ & 2 & 20 & 10 \\
V410 Tau		& W  & 2001 Jan 20		& 6    & 16 & 0.43 & K3 & 4730	& 4 & 4.1 & 7,9 & 83 & 10 \\
V819 Tau		& W  & 2003 Jan 12		& 48   & 15 & 0.43 & K8.0 & 3980  & 1 & 4.0 & 1 & 9 & 10 \\
V827 Tau		& W  & 2003 Jan 13		& 36   & 20 & 0.58 & M1.4 & 3656 & 1 & 3.8 & 1 & 21 & 10 \\
V830 Tau		& W  & 2003 Jan 12		& 48   & 12 & 0.43 & K7.5 & 4000  & 1 & 3.9 & 1 & 32 & 10 \\
\\
\\
TW Hya		& C,T  & 2001 Jan 20	& 12   & 25 & 0.43 & K7 & 4126$\tablenotemark{b}$ & 5 & 4.8$\tablenotemark{b}$ & 5 & 6 &  5 \\
TW Hya		& C,T  & 2003 Jan 11	& 40   & 30 & 0.43 & K7 & 4126$\tablenotemark{b}$ & 5 & 4.8$\tablenotemark{b}$ & 5 & 6 &  5 \\
\\
TWA 2AB		& W  & 2003 Jan 13		& 40   & 30 & 0.58 & M2.2 & 3530 & 1 & 3.9 & 1 & 13 & 12 \\
TWA 5		& W  & 2003 Jan 12		& 26   & 55 & 0.58 & M2.7 & 3455 & 1 & 3.7 & 1 & 55$\tablenotemark{c}$ & 12\\
TWA 6		& W  & 2003 Jan 12		& 32   & 18 & 0.58 & M0.0 & 3900 & 1 & 4.3 & 1 & 55 &  13 \\
TWA 7		& W  & 2003 Jan 12		& 32   & 50  & 0.58 & M3.2 & 3366 & 1 & 3.9 & 1 & 2 & 13 \\
TWA 8A		& W  & 2001 Jan 20		& 16   & 25 & 0.43 & M2.9 & 3425 & 1 & 4.0 & 1 & 8$\tablenotemark{d}$ & 13 \\
TWA 9A		& W  & 2003 Jan 13		& 28   & 10 & 0.58 & K6.0 & 4115 & 1 & 4.5 & 1 & 9 & 14 \\
TWA 10		& W  & 2001 Jan 20		& 8    & 7 & 0.43 & M2.5 & 3525 & 6 & 4.3 & 8,9 & 2 & 8 \\
TWA 10		& W  & 2003 Jan 11		& 44   & 12 & 0.43 & M2.5 & 3525 & 6 & 4.3 & 8,9 & 2 & 8 \\
\\

\enddata
\tablenotetext{a}{Accreting sources (CTTS) with a transition object SED are designated `C,T'. 
All other CTTS are designated `C'.  Non-accreting sources (WTTS) with a transition object SED are designated `W,T'. 
All other WTTS are designated `W'.}
\tablenotetext{b}{$\teff$ and $\logg$ reported directly from spectroscopic measurements}
\tablenotetext{c}{A better model fit was found using 55 instead of 36 $\kms$ from the literature}
\tablenotetext{d}{A better model fit was found using 8 instead of $<3$ $\kms$ from the literature}
\tablecomments{(1) \citet{herczeg2014}, (2) \citet{balachandran1994}, (3) \citet{johns-krull2004}, (4) \citet{kenyon1995}, (5) \citet{yang2005}, (6) \citet{webb1999}, (7) \citet{bertout2007}, (8) \citet{delareza2004}, (9) \citet{baraffe2015}, (10) \citet{nguyen2012}, 
(11) White \& Hillenbrand 2004, (12) \citet{torres2003}, (13) \citet{reid2003}, (14) \citet{yang2008} }
\end{deluxetable}


\begin{deluxetable}{llclc}
\tablecolumns{5}
\tablewidth{0 pt}
\tablecaption{Model Parameters and Derived Properties}
\tablehead{
\colhead{Object}	&
\colhead{$\teff$$\tablenotemark{a}$}	&
\colhead{$\logg$$\tablenotemark{a}$}	&
\colhead{Veiling}	&
\colhead{Radial Velocity$\tablenotemark{b}$}	\\
\colhead{} 	&
\colhead{(K)}	&
\colhead{($\gpersqcm$)}		&
\colhead{($\mveil$)} 	&
\colhead{v$_{\rm helio}$ ($\kms$)}	\\
}

\startdata

DN Tau			& 3800 & 3.5 & 2.5 $(+0.05,-0.7)$ & 16\\
IP Tau			& 3800 & 4.0 & 4.7 $(+0.9,-0.05)$ & 16\\
V836 Tau (2001)	& 3800 & 4.0 & 2.1 $(+0.05,-0.4)$ & 18\\
V836 Tau (2003)	& 3800 & 4.0 & 2.5 $(+0.05,-0.5)$ & 18\\
ZZ Tau			& 3400 & 3.5 & 0.7 $(+0.3,-0.1)$ & 18\\
\\
CoKu Tau/4		& 3700 & 4.0 & 0.3 $(+0.2,-0.05)$ & 16\\
GM Aur			& 4200 & 4.0 & 1.1 $\pm0.1$ & 15\\
LkCa 15 (2001)		& 4200 & 4.0 & 5.1 $(+0.2,-0.3)$ & 18\\
LkCa 15 (2007)		& 4200 & 4.0 & 3.5 $(+0.1,-0.2)$ & 18\\
UX Tau A (2003)	& 5200 & 4.5 & 2.8 $(+0.3,-0.6)$ & 16\\
UX Tau A (2005)	& 5200 & 4.5 & 4.4 $(+0.5,-0.8)$ & 16\\
\\
HBC 427			& 4200 & 4.0 & 0.1 $\pm0.1$ & 17\\
Hubble 4			& 4200 & 3.5 &  0.1 $(+0.3,-0.05)$ & 14\\
IW Tau			& 3700 & 3.5 &  0.3 $(+0.05,-0.1)$ & 16\\
LkCa 19			& 5000 & 4.0 & 0.0 $(+0.05,-0.1)$ & 13\\
V410 Tau			& 4800 & 4.0 & -0.2 $(+0.05,-0.1)$ & 20\\
V819 Tau			& 4000 & 4.0 & 0.1 $\pm0.1$ & 17\\
V827 Tau			& 3700 & 4.0 & 0.3 $(+0.3,-0.05)$ & 18\\
V830 Tau			& 4000 & 4.0 & -0.1 $\pm0.1$ & 18\\
\\
TW Hya (2001)		& 4200 & 4.5 & 1.2 $(+0.4,-0.05)$ & 13\\
TW Hya (2003)		& 4200 & 4.5 & 2.0 $(+0.5,-0.05)$ & 13\\
\\
TWA 2AB			& 3500 & 4.0 & 0.4 $(+0.2,-0.05)$ & 11\\
TWA 5			& 3500 & 3.5 & 0.4 $(+0.2,-0.1)$ & 15\\
TWA 6			& 3900 & 4.5 & 0.2 $(+0.2,-0.05)$ & 17\\
TWA 7			& 3400 & 4.0 & 0.3 $(+0.2,-0.05$ & 12\\
TWA 8A			& 3400 & 4.0 & 0.1 $\pm0.2$ & 8\\
TWA 9A			& 4200 & 4.5 & 0.0 $(+0.05,-0.1)$ & 10\\
TWA 10 (2001)		& 3500 & 4.5 & -0.1 $(+0.2,-0.05)$ & 7\\
TWA 10 (2003)		& 3500 & 4.5 & 0.0 $(+0.2,-0.05)$ & 7\\

\enddata
\tablenotetext{a}{Used stellar atmosphere model that was closest to literature values reported in Table 1}
\tablenotetext{b}{Measured radial velocities are in agreement with literature values except for the following spectroscopic binaries: HBC 427, Hubble 4, and TWA 5}

\end{deluxetable}

%

\begin{deluxetable}{ l l l l l c l l l l }
\tablecolumns{9}
\tablewidth{0 pt}
\tablecaption{Measured CO Emission Properties}
\tablehead{
\colhead{Object}				&
\multicolumn{4}{c}{CO (1-0) low J}	& &
\multicolumn{4}{c}{CO (1-0) high J}	\\ \cline{2-5} \cline{7-10}
\colhead{}						&
\colhead{EW$\tablenotemark{b}$}					&
\colhead{error$\tablenotemark{b}$}					&
\colhead{FWHM}				&
\colhead{J lines used}			& &
\colhead{EW$\tablenotemark{b}$}					&
\colhead{error$\tablenotemark{b}$}					&
\colhead{FWHM}				&
\colhead{J lines used}			\\
\colhead{}						&
\colhead{(\AA)}					&
\colhead{}					&
\colhead{($\kms$)}				& & 
\colhead{}						&
\colhead{(\AA)}					&
\colhead{}					&
\colhead{($\kms$)}				&
\colhead{}					 
}
\startdata
DN Tau$\tablenotemark{a}$		& 2.26 & $\pm0.03$ & 128 & R1-3 && 3.37  & $\pm0.07$ & 130 & P26,P30  \\
IP Tau$\tablenotemark{a}$		& 2.18 & $\pm0.04$ & 108 & R3-5 && 3.42  & $\pm0.05$ & 120-150 & P25-P28  \\
V836 Tau (2001)$\tablenotemark{a}$ 	& 3.88 & $\pm0.12$ & 125 & R2,R3,R5 && 4.65  & $\pm0.24$ & 85 & P27,P28,P30 \\
V836 Tau (2003)$\tablenotemark{a}$ 	& 3.19 & $\pm0.05$ & 128 & R1,R3,R5 && 3.24  & $\pm0.20$ & 120 & P26,P28,P30  \\
ZZ Tau			& $<0.30 $ &- & - & R3 && $<0.35$ & - & - & P26 \\
\\
CoKu~Tau/4		& $<0.29$ &- & - &  R3 && $<0.34$ & - & - & P26 \\
GM Aur			& 0.33 & $\pm0.01$ & 22 & R2,R3,R5 && 0.56  &  $\pm0.01$ & 30 & P26-27  \\
LkCa 15 (2001)		& 0.34 & $\pm0.01$ & $\sim$20 & R2,R3,R5 && 1.68  &  $\pm0.04$ & 73 & P25-28 \\
LkCa 15 (2007)		& $<0.29$ & -  & - & R3 && 0.2  &  $\pm0.05$ & 30 & P26-27,P30  \\
UX Tau A (2003)		& 0.48 & $\pm0.03$ & 30 & R0-3 && $<0.24$  & - &  - & P26 \\
UX Tau A (2005)		& 0.31 & $\pm0.01$ & 22 & R0-3 && 0.19  &  $\pm0.01$ & 28 & P25-26,P28-29 \\
\\
HBC~427			& $<0.31$ &- & - &  R3 && $<0.33$ & - & - & P26 \\
Hubble 4		& $<0.16$ &- & - &  R3 && $<0.18$ & - & - & P26 \\
IW Tau			& $<0.34$ &- & - &  R3 && $<0.28$ & - & - & P26 \\
LkCa~19			& $<0.34$ &- & - &  R3 && $<0.29$ & - & - & P26 \\
V410~Tau		& $<0.34$ &- & - &  R3 && $<0.33$ & - & - & P26 \\
V819~Tau		& $<0.28$ &- & - &  R3 && $<0.28$ & - & - & P26 \\
V827~Tau		& $<0.24$ &- & - &  R3 && $<0.26$ & - & - & P26 \\
V830~Tau		& $<0.31$ &- & - &  R3 && $<0.36$ & - & - & P26 \\
\\
TW Hya (2001)		& $<0.36$ &- & - & R3 && 1.10  & $\pm0.03$ & 16 & P26-27 \\
TW Hya (2003)		& $<0.43$ &- & - & R3 && 1.28  & $\pm0.05$ & 15 & P26-27 \\
\\
TWA~2AB			& $<0.23$ &- & - &  R3 && $<0.24$ & - & - & P26 \\
TWA~5			& $<0.13$ &- & - &  R3 && $<0.16$ & - & - & P26 \\
TWA~6			& $<0.40$ &- & - &  R3 && $<0.34$ & - & - & P26 \\
TWA~7			& $<0.31$ &- & - &  R3 && $<0.30$ & - & - & P26 \\
TWA~8A			& $<0.31$ &- & - &  R3 && $<0.32$ & - & - & P26 \\
TWA~9A			& $<0.51$ &- & - &  R3 && $<0.38$ & - & - & P26 \\
TWA~10~(2001)	& $<1.06$ &- & - &  R3 && $<0.53$ & - & - & P26 \\
TWA~10~(2003)	& $<0.62$ &- & - &  R3 && $<0.39$ & - & - & P26 \\


\enddata

\tablenotetext{a}{Averaged low J lines measured from half width of emission profiles and assuming symmetry}
\tablenotetext{b}{Equivalent widths and errors are measured relative to the observed continuum level}\end{deluxetable}

\end{document}